\documentclass[useAMS,usenatbib,usegraphicx]{pasj00}
\usepackage{times}
\usepackage{natbib}
\usepackage{graphicx}
\usepackage{fixltx2e}
\usepackage{mathptmx}

\def\c{{\em Chandra}}
\def\jwst{{\sl JWST}}

\def\hst{{\sl HST}}

\def\einstein{{\sl Einstein}}
\def\rosat{{\em ROSAT}}

\def\spica{{\em Spica}}
\def\suzaku{{\em Suzaku}}

\def\iras{{\em IRAS}}

\def\p{$\pm$}
\def\ltsim{\mathrel{\hbox{\rlap{\hbox{\lower4pt\hbox{$\sim$}}}\hbox{$<$}}}}
\def\gtsim{\mathrel{\hbox{\rlap{\hbox{\lower4pt\hbox{$\sim$}}}\hbox{$>$}}}}

\def\Msun{M$_{\odot}$}
\def\lsun{$L_{\odot}$} 
\def\micron{$\mu$m}
\def\araa{ARA\&A}
\def\aap{A\&A}
\def\nat{Nat}
\def\aaps{A\&AS}
\def\mnras{MNRAS}
\def\apj{ApJ}
\def\apjs{ApJS}
\def\aj{AJ}
\def\apjl{ApJL}
\def\pasp{PASP}
\def\pasj{PASJ}

\def\nh{$N_{\rm H}$}

\def\iraf{{\sc iraf}}
\def\xspec{{\sc xspec}}
\def\ciao{{\sc ciao}}
\def\caldb{{\sc caldb}}

\def\hii{H{\sc ii}}

\def\l{$\lambda$}

\def\neii{[Ne{\sc ii}]}

\def\av{$A_{\rm V}$}
\def\n117filter{N11.7}
\def\neiifilter{NeII}
\Received{$\langle$2010 Nov 24$\rangle$}
\Accepted{$\langle$2011 Jan 17$\rangle$}
\SetRunningHead{Astronomical Society of Japan}{Subaru mid-infrared imaging of the core of M82}
\begin{document}

\title{Diffraction-limited Subaru imaging of M82: sharp mid-infrared view of the starburst core\thanks{Based in part on data collected at Subaru Telescope, which is operated by the National Astronomical Observatory of Japan.}}

\author{P. Gandhi}
\affil{Institute of Space and Astronautical Science (ISAS), Japan Aerospace Exploration Agency, 3-1-1 Yoshinodai, chuo-ku, Sagamihara, Kanagawa 252-5210, Japan}
\email{pgandhi@astro.isas.jaxa.jp}
\author{N. Isobe}
\affil{Department of Astronomy, Kyoto University, Kitashirakawa-Oiwake-cho, Sakyo-ku, Kyoto 606-8502, Japan
}
\author{M. Birkinshaw, D.M. Worrall}
\affil{H.H. Wills Physics Laboratory, University of Bristol, Tyndall Ave, Bristol BS8 1TL, UK
}
\author{I. Sakon}
\affil{Department of Physics, University of Tokyo, 7-3-1 Hongo, Bunkyo-ku, Tokyo 113-0033, Japan}
\author{K. Iwasawa}
\affil{ICREA and Institut de Ci\`encies del Cosmos, Universitat de Barcelona, Mart\'i i Franqu\`es, 1, 08028 Barcelona, Spain}
\author{A. Bamba}
\affil{School of Cosmic Physics, Dublin Institute for Advanced Studies 31 Fitzwilliam Place, Dublin 2, Republic of Ireland}
\affil{Institute of Space and Astronautical Science (ISAS), Japan Aerospace Exploration Agency, 3-1-1 Yoshinodai, chuo-ku, Sagamihara, Kanagawa 252-5210, Japan}

\maketitle

\begin{abstract}We present new imaging at 12.81 and 11.7 $\mu$m of the central $\sim$40$''$$\times$30$''$ ($\sim$0.7$\times$0.5 kpc) of the starburst galaxy M82. The observations were carried out with the COMICS mid-infrared (mid-IR) imager on the 8.2~m Subaru telescope, and are diffraction-limited at an angular resolution of $<$0$\farcs$4. The images show extensive diffuse structures, including a 7\arcsec--long linear chimney-like feature and another resembling the edges of a ruptured bubble. This is the clearest view to date of the base of the kpc-scale dusty wind known in this galaxy. These structures do not extrapolate to a single central point, implying multiple ejection sites for the dust. In general, the distribution of dust probed in the mid-IR anticorrelates with the locations of massive star clusters that appear in the near-infrared. The 10--21~\micron\ mid-IR emission, spatially-integrated over the field of view, may be represented by hot dust with temperature of $\sim$160 K. Most discrete sources are found to have extended morphologies. Several radio \hii\ regions are identified for the first time in the mid-IR. The only potential radio supernova remnant to have a mid-IR counterpart is a source which has previously also been suggested to be a weak active galactic nucleus. This source has an X-ray counterpart in \c\ data which appears prominently above 3 keV and is best described as a hot ($\sim$2.6 keV) absorbed thermal plasma with a 6.7 keV Fe K emission line, in addition to a weaker and cooler thermal component. The mid-IR detection is consistent with the presence of strong \neii\l 12.81\micron\ line emission. The broad-band source properties are complex, but the X-ray spectra do not support the active galactic nucleus hypothesis. We discuss possible interpretations regarding the nature of this source.
\end{abstract}

\section{Introduction}

The galaxy M82 (NGC~3034) hosts the nearest and best example of an ongoing massive starburst, making it an excellent target for detailed studies at all wavelengths. The galaxy is thought to have undergone an interaction event with its neighbor M81 about 10$^8$ yr ago \citep{gottesman77}, triggering a massive nuclear starburst about 5$\times$10$^7$~years ago \citep{rieke80}. There is also evidence for several other star formation episodes, both older and younger \citep{degrijs01, forsterschreiber03}. Around 40 supernova remnants (SNRs) have been identified in the core \citep{fenech08}, and one new supernova (SN) is produced every $\sim$3 years \citep[e.g. ][]{rieke80, jones84}. The energy output of the super star clusters hosting the SNe is thermalised and drives a large scale superwind along the galactic minor axis \citep{heckman90} which can be observed in detail because of the favorable edge-on inclination of the source. 

In the infrared, M82 has been extensively studied with all space missions. Its infrared luminosity is measured to be 5$\times$10$^{10}$ $L_{\odot}$, and it shows prodigious dusty outflows and polycyclic aromatic hydrocarbon (PAH) grains in the mid-infrared (mid-IR) extending on kpc-scales \citep[e.g. ][]{iraspsc88, sturm00, forsterschreiber03b, engelbracht06, kaneda10, roussel10}. There are very few high-spatial-resolution mid-IR studies of the core itself on scales of order 100 pc, because of size limitations of space missions. With their large primary mirrors, ground-based telescopes can achieve the best spatial resolution currently possible. In the $N$ (8--13 \micron) band atmospheric window, the highest resolution studies thus far are the works of \citet[][hereafter TG92]{telesco92} and \citet[][ hereafter AL95]{achtermannlacy95}, with nominal resolutions of 1$\farcs$1 and 2\arcsec, respectively. 

In this work, we present the first sub-arcsec mid-IR $N$-band images of the core of M82. The galaxy was observed as an extension of our recent work of Seyfert galaxies \citep{g09_mirxray}, as part of a study to understand the mid-IR emission of galaxies at high angular resolution. The observations were carried out at the 8.2~m Subaru telescope. At wavelengths of 11.7 and 12.81 \micron, our imaging is diffraction-limited at $<0\farcs 4$. These new images provide the sharpest mid-IR view of structures at the base of the superwind, and allow an extensive multi-wavelength comparison of individual sources. Several \hii\ regions are newly identified in the mid-IR. We also discuss the nature of a putative active galactic nucleus candidate. Using a mid-IR detection and new \c\ data, we rule out the AGN hypothesis and discuss other possibilities, include supernova remnant ionization and emission from a starburst. 

Distance estimates to M82 have ranged over 3.2--5.2 Mpc \citep[e.g. ][]{burbidge64,tully88,sakaimadore99}. Some of the latest measurements suggest a distance at the lower end of this range based upon accurate determination of the tip of the red giant branch magnitude \citep{karachentsev04,dalcanton09}. We adopt a value of 3.53 Mpc herein, resulting in a physical scale of 17.1 pc per arcsec. Our imaging resolution limit corresponds to $\approx$6.1 and 6.7 pc at 11.7~\micron\ and 12.8~\micron, respectively.

\section{Observations}
\label{sec:obs}

Observations were carried out at Subaru on the night starting 2009 May 04, under generally clear weather conditions. The source was observed at the beginning of the night close to meridian, i.e. near its maximum altitude of 40$^\circ$. The Cooled Mid-Infrared Camera and Spectrometer, or COMICS \citep{comics} was used in imaging mode with standard chopping and nodding off-source. Due to the extended nature of the emission from M82, a relatively large chop throw of 30\arcsec\ was used so that the sky position lay completely outside a single field-of-view of the detector. Technical difficulties prevented us from using larger throws. A North-South chop direction was adopted, because this is approximately perpendicular to the apparent major axis of the galaxy. Integration times of 60 s were used at each chop position. Subsequently, the telescope was nodded to a position at an offset of 1 arcmin to the North, and the above chopping series was replicated, this time entirely on sky.

The above sequence was repeated 10 times, resulting in a total on-source exposure of 600 s per filter and a total telescope time of about 50 min per filter after accounting for observing efficiency. Imaging was obtained in the \neiifilter\ (narrow-band) and the \n117filter\ filters, with central wavelengths (and full widths at half maximum) of 12.81 (0.2) \micron\ and 11.7 (1.0) \micron, respectively\footnote{http://canadia.ir.isas.jaxa.jp/comics/open/guide/filter}. The median airmass of the source was 1.56 and 1.65, respectively.

The Cohen standard star PI2~UMa (HD~73108; \citealt{cohen99}) was observed for photometric calibration just before the target at an airmass of $\approx$1.44. On-chip chopping with a throw of 10\arcsec\ was used, and two nod observations were obtained with exposure time 10 s each. Seeing was measured in this observation, as well as in a standard star observation of HD~108381 (with identical observational setup to PI2~UMa; in the \n117filter\ filter only) carried out approximately 1 h after the target observation was completed. The maximum systematic error in the photometry due to the absolute calibration of the standard star is $\approx$3\%, and this error is added quadratically to all target photometric errors.

\section{Data reduction}

Data reduction was carried out using routines provided by the COMICS team ({\tt q\_series} package\footnote{http://www.naoj.org/Observing/DataReduction/}), and following the procedures recommended in the COMICS Data Reduction Manual v.2.1.1. Some additional image manipulation was carried out in \iraf\footnote{http://iraf.noao.edu/}. The COMICS data flow returns two sets of images for each observation: 1) a \lq {\sc coma}\rq\ cube of data frames including all images obtained in one observing block; 2) a single chop-subtracted \lq {\sc comq}\rq\ image, co-added over all the frames in that block. 

The recommended flat fielding procedure for imaging is a \lq self-sky-flat\rq, in which the \lq off\rq\ beam chop positions are used for flat fielding the \lq on\rq\ beam ones, and vice-versa. This is possible because the detector is illuminated quite uniformly by sky background, much brighter than the imaged sources. To create these flats, dark frames were first subtracted from each cube of data. For the standard star reduction, the on and the off chop beam data cubes were averaged to create a separate mean image for each of the two beams (hereafter, called \lq mean beam images\rq). Low spatial frequency trends across the detector are then removed by simply dividing through a heavily Gaussian-smoothed version of the mean beam images. This results in the desired flats for one observing block, and each {\sc comq} co-added frame can be divided by these flats to create calibrated on and off beam standard star images, with the pixel-to-pixel sensitivity variation removed. This flat fielding procedure was carried out separately for each nod position, following which the two beams are shifted and averaged (with flux inversion of one). 

The target was observed with a large chop angle so that the off beam images are devoid of any bright sources. Hence, only the flats created from the off beam positions are required to calibrate the on beam target images. So, for the target, we simply divided each {\sc comq} image by the off beam flat created as described above. This was done identically for the nod position as well. Each flat fielded nod position image was then subtracted from its immediately preceding flat fielded target image, resulting in 10 calibrated and background-subtracted images of the target. One of the point-like sources in each image was used as a reference for determining small shifts necessary to create the final co-added image. 

\begin{figure*}
  \begin{center}
    \includegraphics[width=15cm]{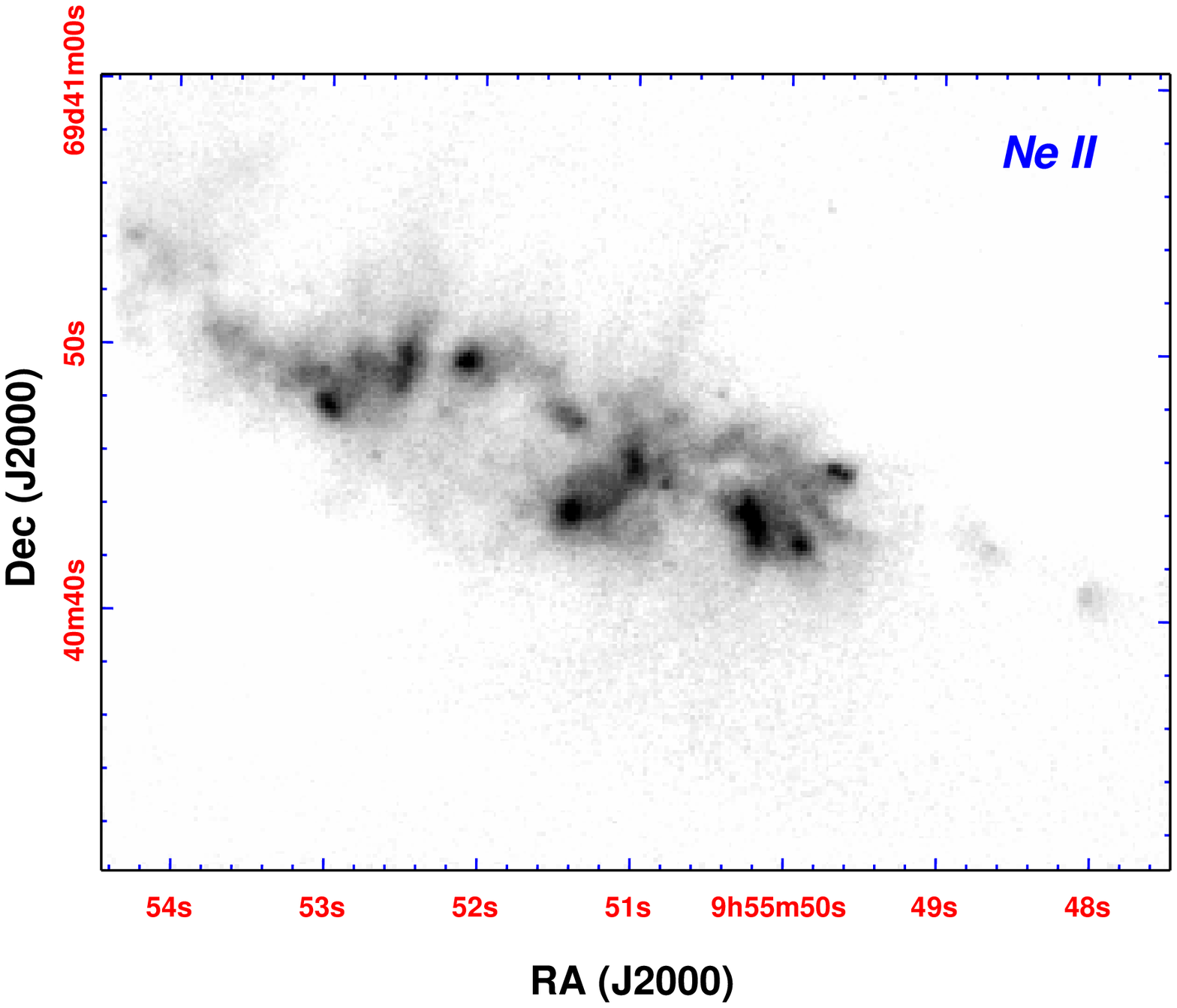}
    \includegraphics[width=1.66cm]{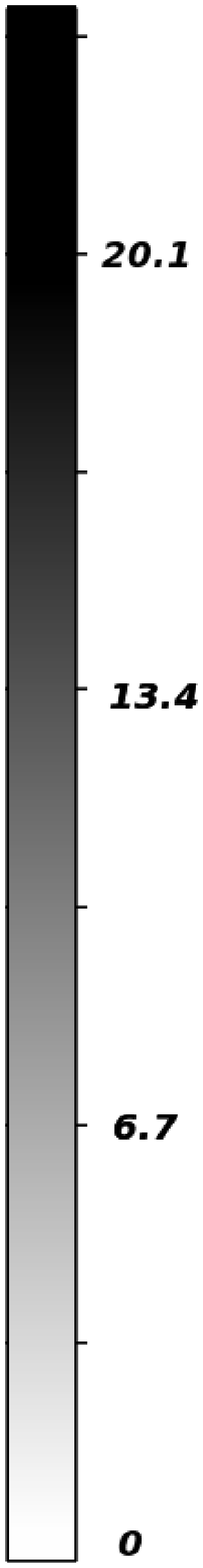}
    \includegraphics[width=15cm]{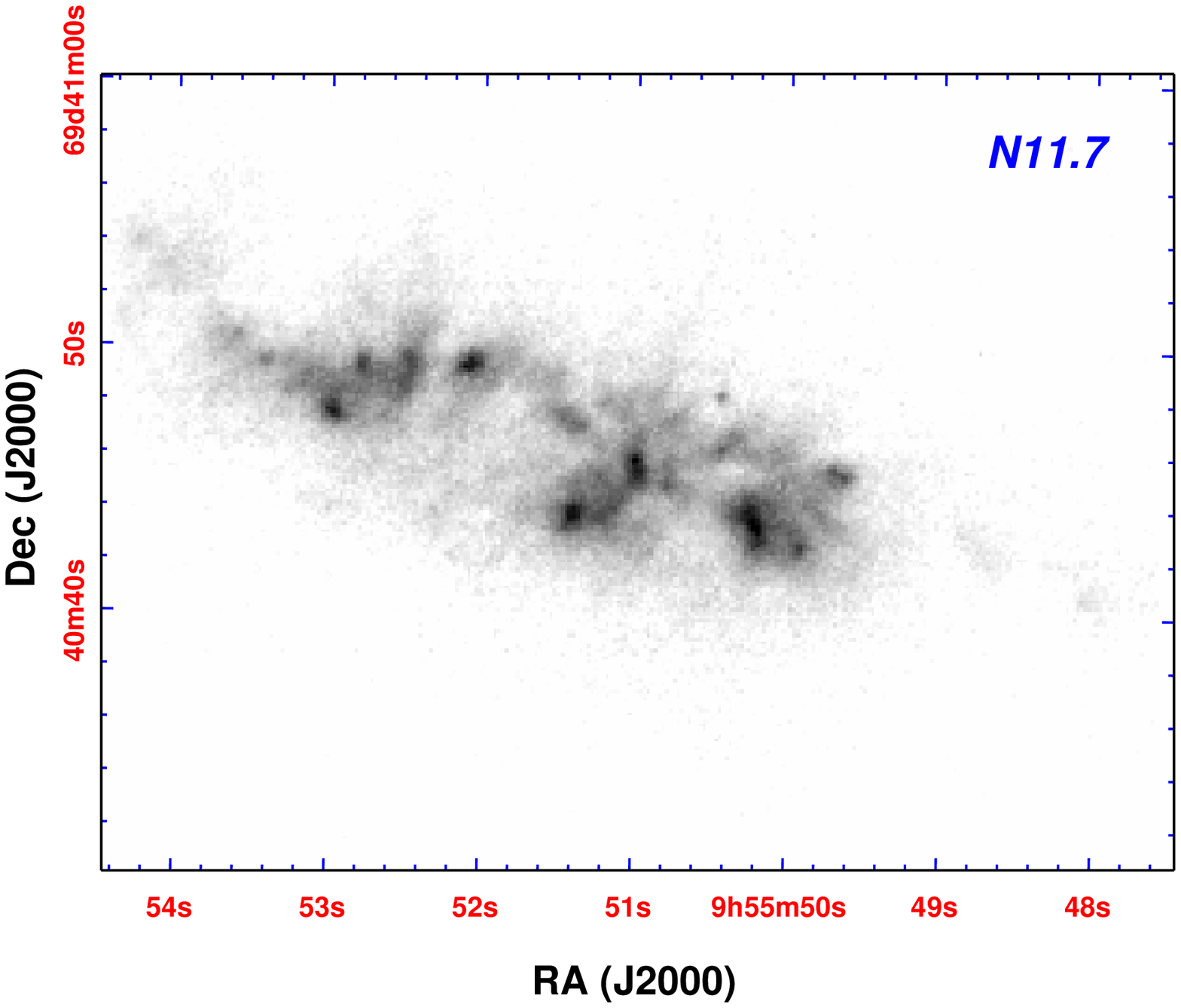}
    \includegraphics[width=1.66cm]{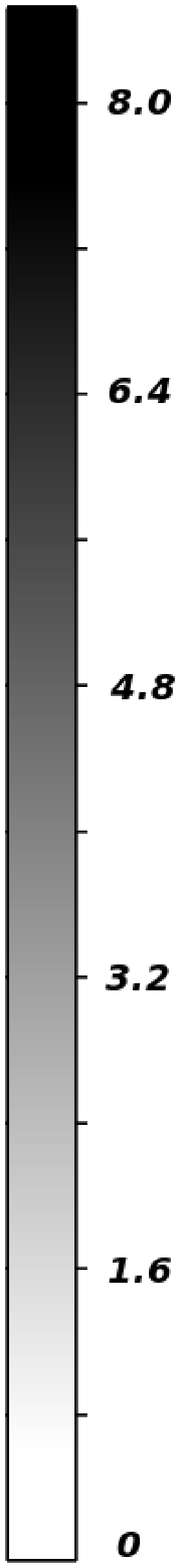}
 \end{center}
  \caption{COMICS images in the \neiifilter\ filter (top) and \n117filter\ filter (bottom). The field of view displayed is 286x213 pixels wide which, at a pixel scale of 0$\farcs$133, results in a field size of 38\arcsec $\times$28\arcsec. The linear grayscale is in units of mJy pix$^{-1}$. 
}\label{fig:basicimages}
\end{figure*}

Two systematic sources of noise affect COMICS observations. Firstly, a low level random noise in the form of horizontal stripes is introduced into every exposure by the readout amplifiers. Although it is largely periodic in nature across the field of view as a result of the 16 identical amplifier channels employed, this is difficult to remove because of 1) its random nature from exposure to exposure; 2) the fact that our target covers most of the field of view, leaving free very little of the CCD for determination of the noise level. No optimal solution was found for removal of this read noise component without introducing additional statistical uncertainties in the process, so no attempt is made to remove it, except for display purpose.

Secondly, we find non-negligible background residuals in our \n117filter\ image. This is a result of the fact that 1) the chop throw (and nod offset) is relatively large which means that the background and target optical light paths differ significantly; and 2) the typical time delay between a chopped beam observation and its corresponding nod observation is of order several minutes, during which time the sky level can vary significantly. These residuals are more prominent in the \n117filter\ image because of the wide filter width could make it more sensitive to sky variations. In order to remove these, the final NeII image was rescaled in order to produce the best match to the target in the \n117filter\ filter, and then subtracted. This leaves a background image dominated by the \n117filter\ background residuals, with only some additional low level structures due to the varying SEDs of targets across the field of view between the two wavelengths. This background image can then be subtracted from the shifted and co-added \n117filter\ image to create the desired residual-free product.

\vspace*{0.5cm}
\noindent
Final calibrated images are shown in Fig.~\ref{fig:basicimages}. The attached absolute astrometry was determined based on multi-wavelength comparison of detected sources, as described in the following sections. The coordinates attached to all images presented herein are for the J2000 equinox.

\section{Point spread function}

\begin{figure*}
  \begin{center}
    \includegraphics[width=5.5cm]{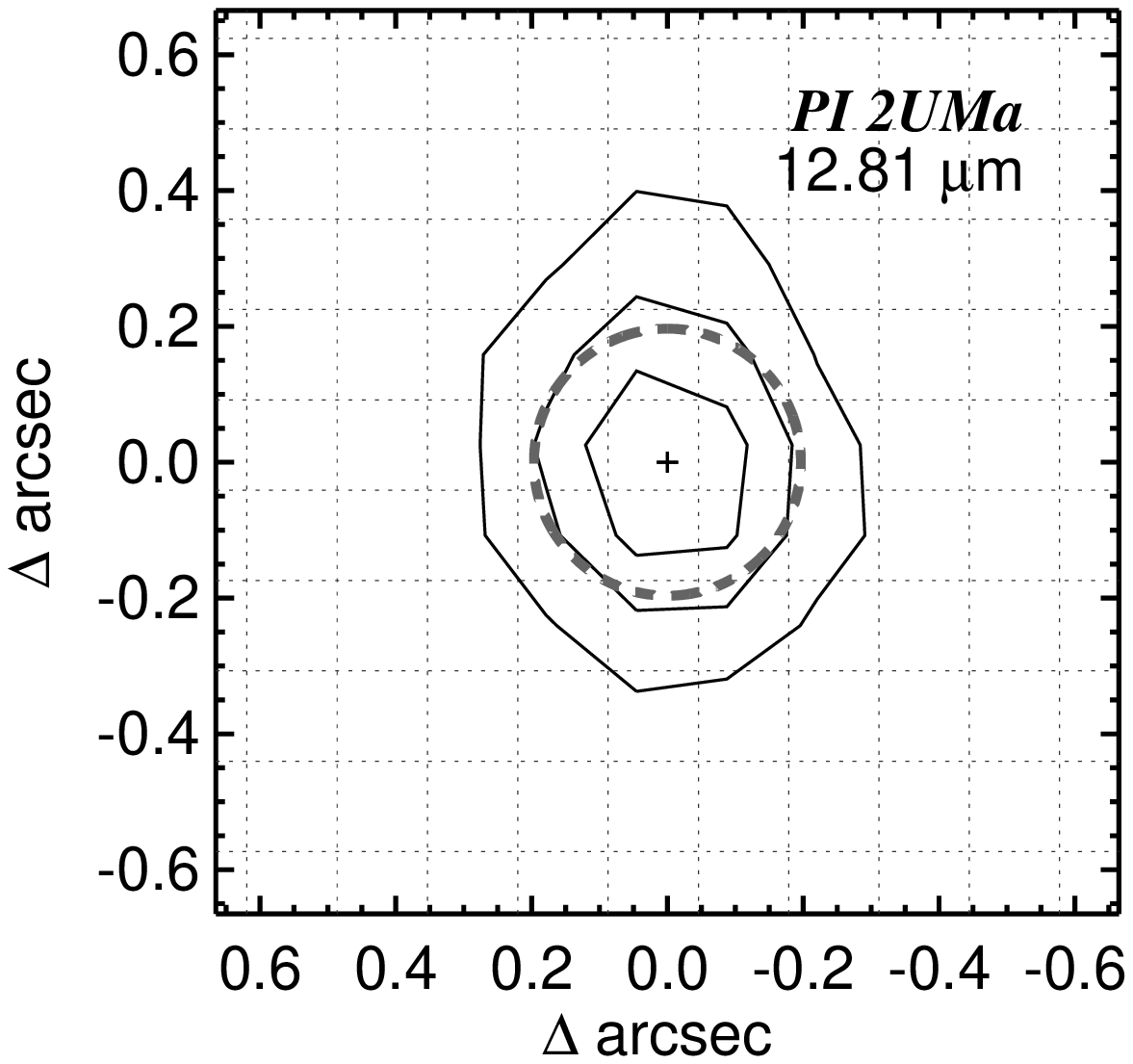}
    \includegraphics[width=5.5cm]{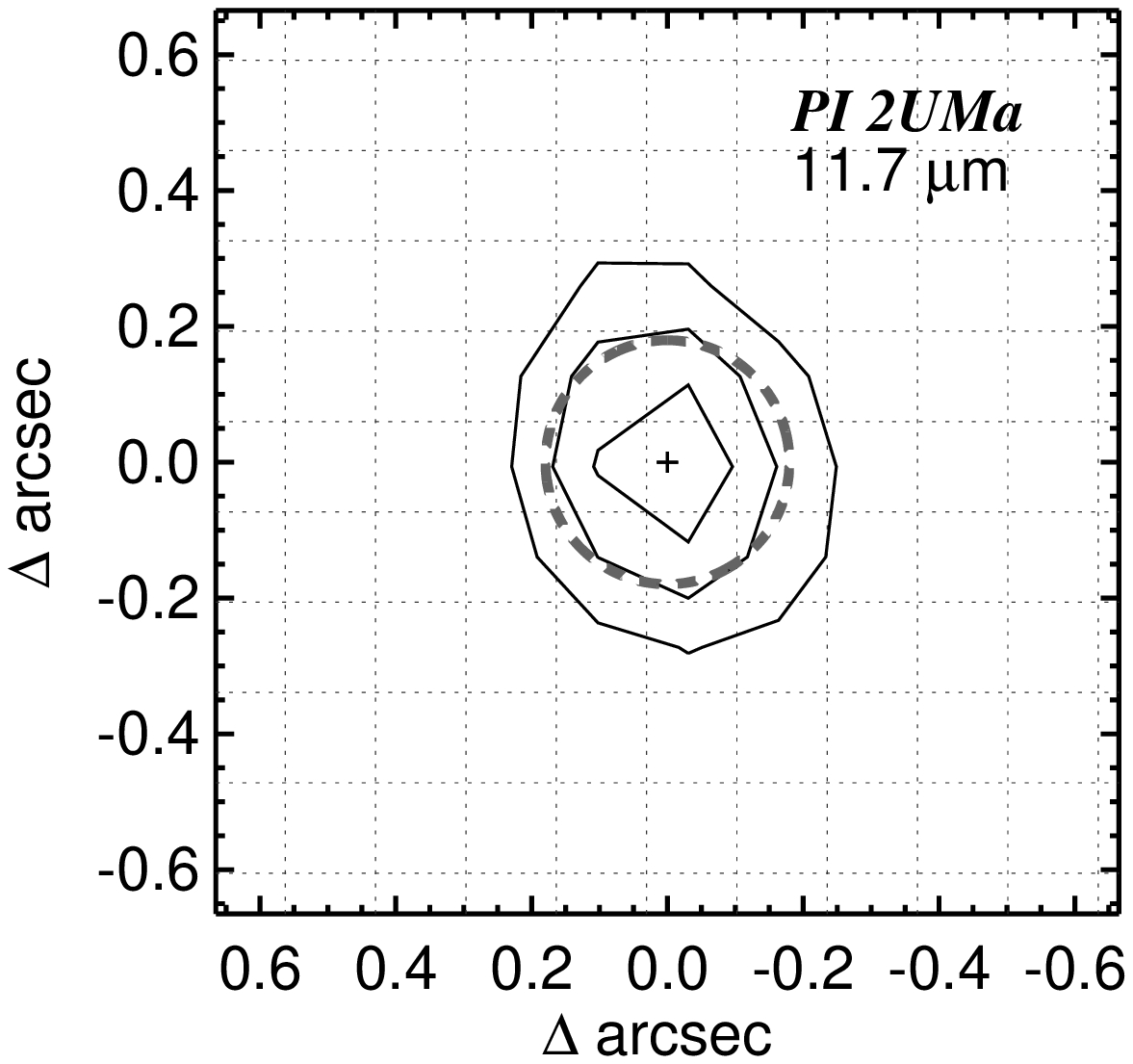}
    \includegraphics[width=5.5cm]{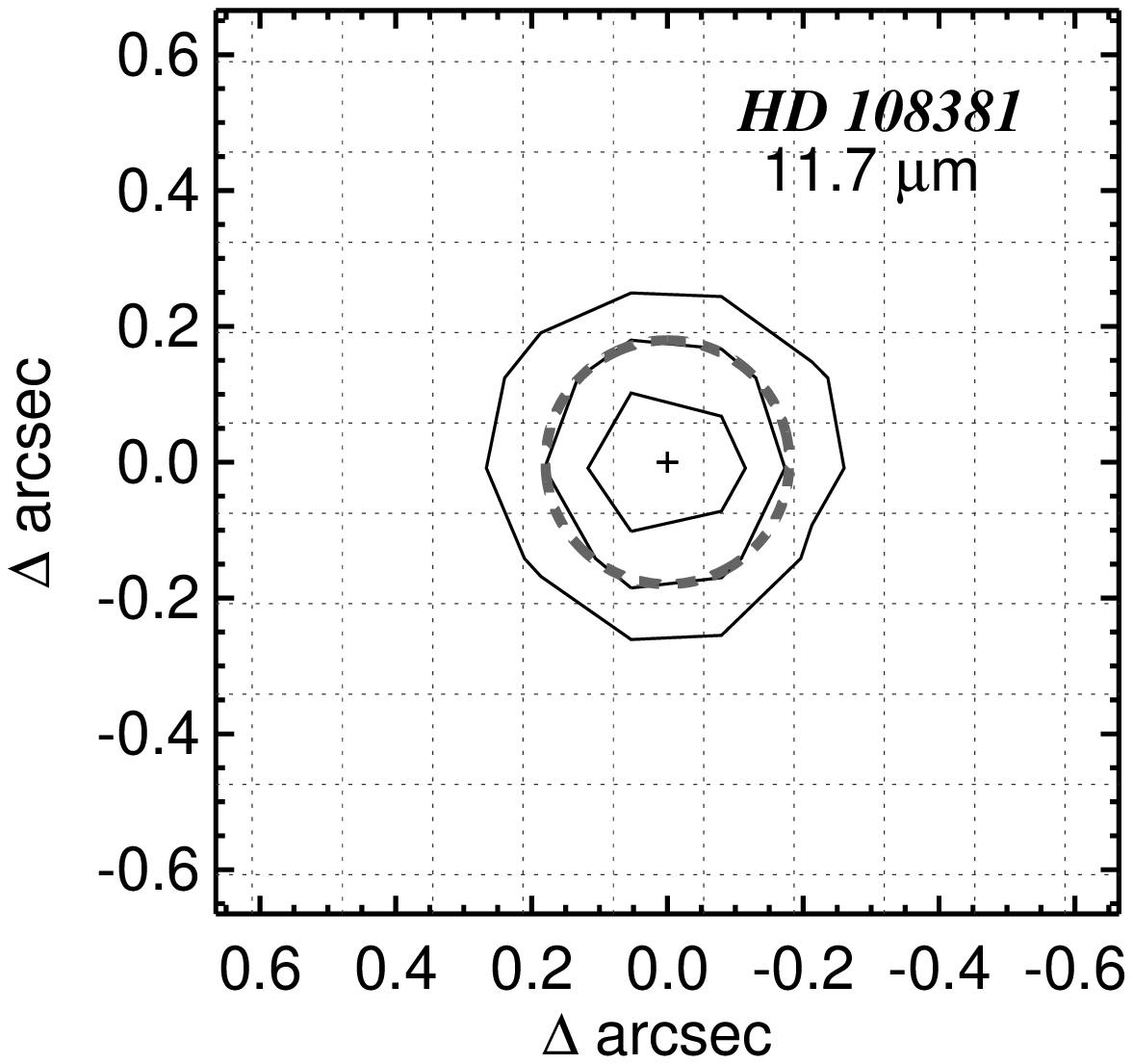}
  \end{center}
  \caption{PSF measurement from standard stars observed immediately prior to (PI 2UMa), and about one hour after (HD 108381), the target observation. The filter central wavelengths ($\lambda$) are labeled. Contour levels are drawn at 25, 50, 75 and 100 \% (last one is a cross) of peak counts. The thick dashed gray curve is a circle with diameter 1.22$\lambda$/$d_M$ (where $d_M$ is the primary mirror diameter of 8.2 m) and is seen to match well with the 50 \% contour corresponding to the full-width at half maximum. The dotted grid denotes the array pixel scaling of 0\farcs 133. This shows that the imaging is closely diffraction-limited.}\label{fig:psfs}
\end{figure*}

The point-spread-function (PSF) size in our observations may be accurately measured from the photometric standard star observations, which were carried out both before, and after, the target observations. The stellar flux contours are displayed in Fig.~\ref{fig:psfs}. Also overplotted is the expected diffraction limit resolution for both filters. The 50 per cent stellar flux contour matches the expected ideal full-width at half-maximum (FWHM), meaning that our observations were closely diffraction-limited at 0\farcs 36 and 0\farcs 39 in the \n117filter\ and NeII filters, respectively.

\section{Source detection and photometry}
\label{sec:photometry}

In order to isolate the most significant discrete sources in the target field of view, we carried out automatic detection with the SExtractor package \citep{sextractor}. The dominant source of uncertainty for source detection and photometry is the diffuse and bright galactic background emission. An iterative process of interpolating the image on a smooth mesh and removing detected sources is employed for proper modeling of this background. But the complex nature of the observed emission makes the local background quite sensitive to the choice of the mesh size. We thus decided to run SExtractor in two passes using different mesh sizes. In the first instance, a background map is created from a mesh with squares of size 8$\times$8 pixels. A limit of 3$\sigma$ (with $\sigma$ being the root-mean-square variance in the background map at the mesh location of interest) was chosen for defining significant sources, and a rather small minimum detection area of 3 contiguous pixels with a significant signal was adopted for detecting the more compact sources. In a second pass, we chose a smaller background mesh size of 4$\times$4 pixels, and ran the source detection procedure again. SExtractor fits ellipse profiles to all sources, and only those sources whose centroids remained consistent to within some threshold between the two passes were retained in the end. A threshold of size equal to the semi-minor axis for the \neiifilter\ filter (and twice this value for the \n117filter\ images which are affected by worse residuals, as described above) was found from experimentation to work well. Note that a very small background mesh can result in subtraction of some flux from the source wings themselves. Test runs on the photometric star images showed that using a mesh as small as 4$\times$4 pixels results in fluxes lower than the total expected values by factors of about 1.2 and 1.3 (\n117filter\ and \neiifilter, respectively), so all SExtractor--determined fluxes have been increased by these amounts. Because of the clumpy nature of the galaxy emission, additional systematic uncertainties of $\approx$10\%\ are estimated for this background correction, and these have also been included. A better photometric solution will require accurate mapping of the galactic background emission (in longer exposures or using space telescopes), which the present data do not allow. 

\begin{figure*}
  \begin{center}
    \includegraphics[width=15cm]{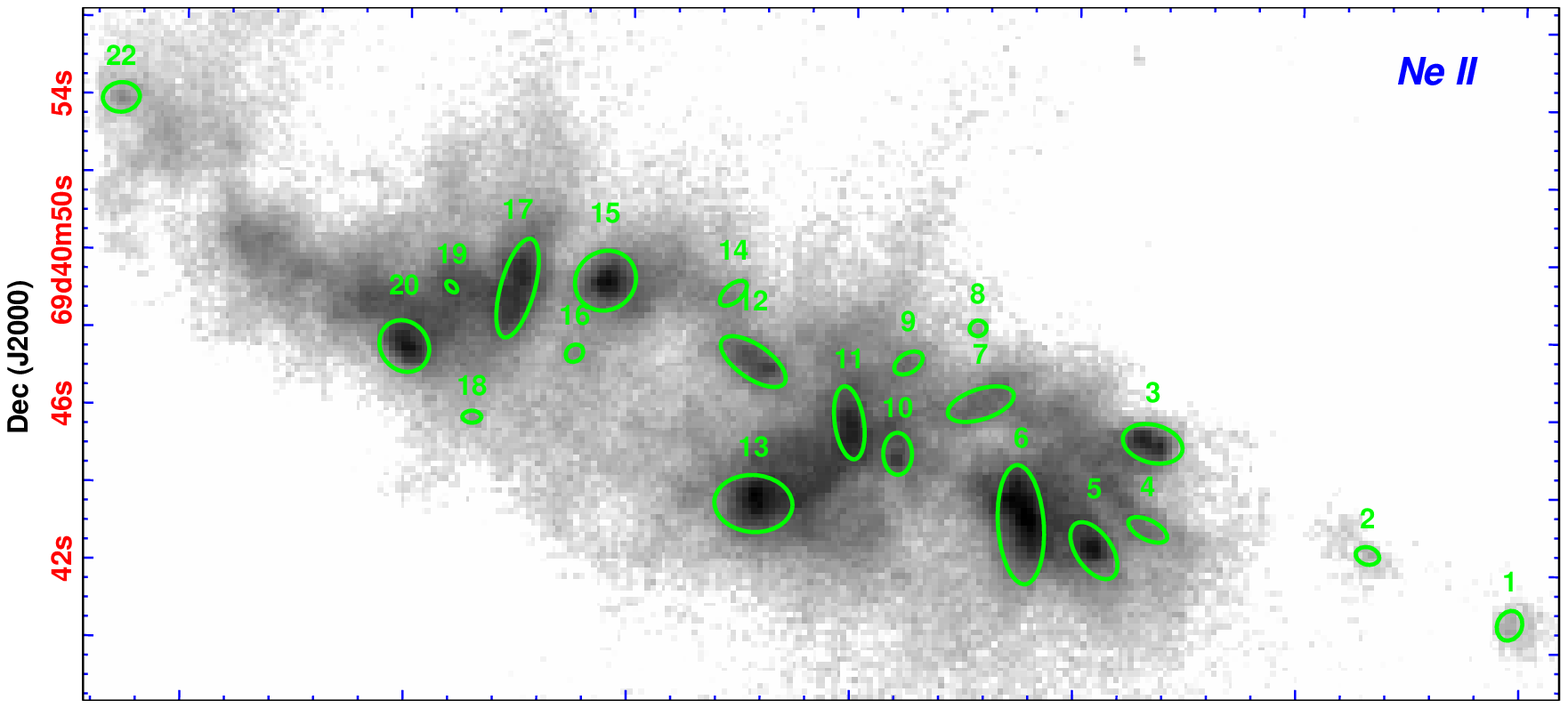}
    \includegraphics[width=15cm]{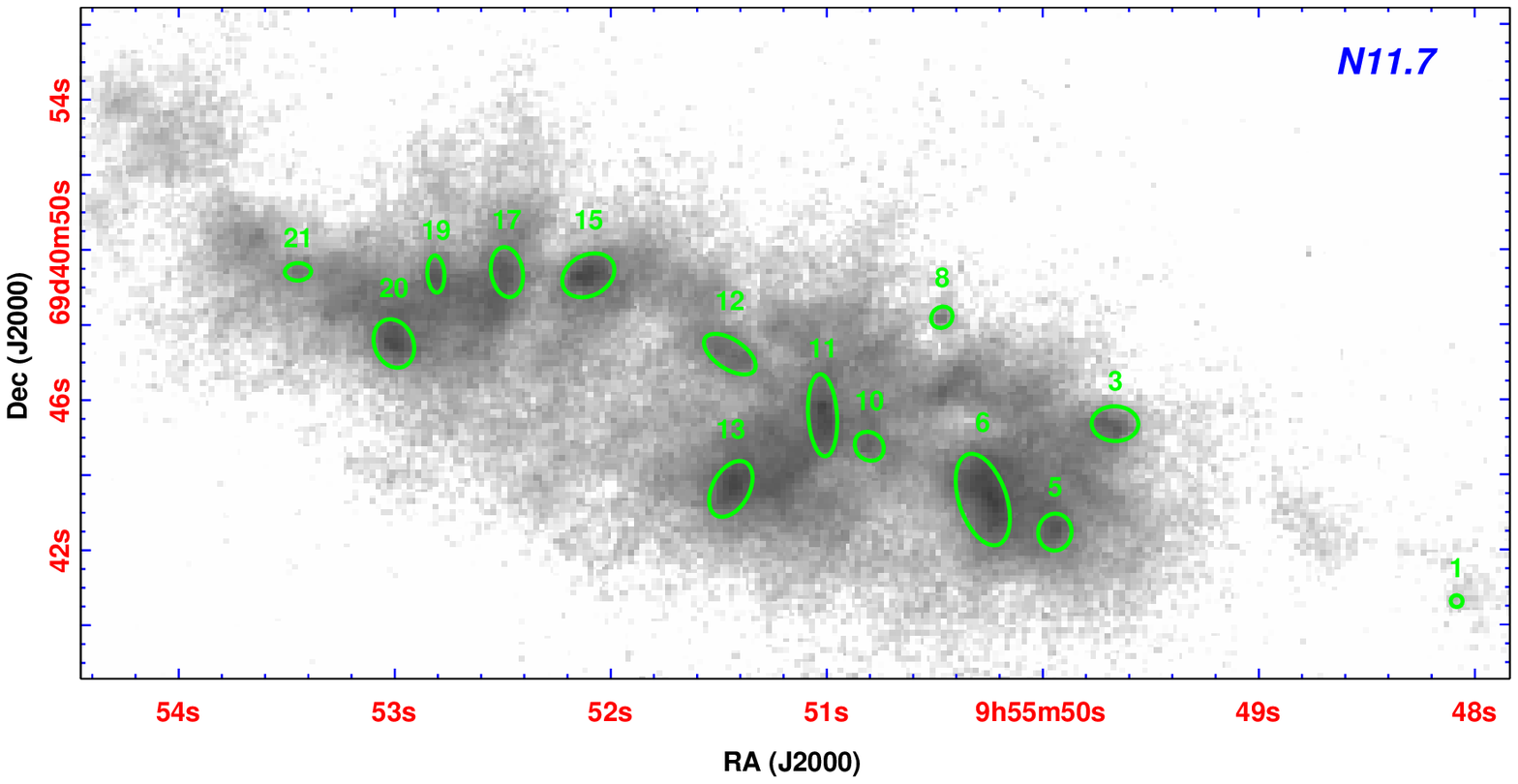}
  \end{center}
  \caption{Detected discrete sources in the NeII {\em (Top)} and N117 {\em (Bottom)} filters. Numeric identifications as in Table~\ref{tab:compact}. The source position angle and extents are those determined from isophotal photometry with elliptical profiles by SExtractor.}\label{fig:compact}
\end{figure*}

Calibration of counts to flux was carried out using the photometric standard observations described in \S~\ref{sec:obs}. Final source fluxes are listed in Table~\ref{tab:compact} and the sources are identified in Fig.~\ref{fig:compact}. 

Finally, in order to determine limiting fluxes for non detections, we simulated two dimensional Gaussian profiles matched to the PSF, scaled these according to flux and added them in at specific positions in the images where limits were required. We then passed these images through our detection pipeline, and determined the flux limit corresponding to the faintest source that could be detected. We find nominal point source flux limits of $\approx$43 mJy and 18 mJy in the \neiifilter\ and \n117filter\ filters, respectively, at the radio kinematic center position of \citet{weliachew84}. But these limits can vary significantly with position and source morphology assumed, so we quote relevant numbers for some interesting sources individually in the following sections. 

\section{Multi-wavelength registration of images}
\label{sec:image_registration}

\begin{figure*}
  \begin{center}
    \includegraphics[width=8.9cm,angle=0]{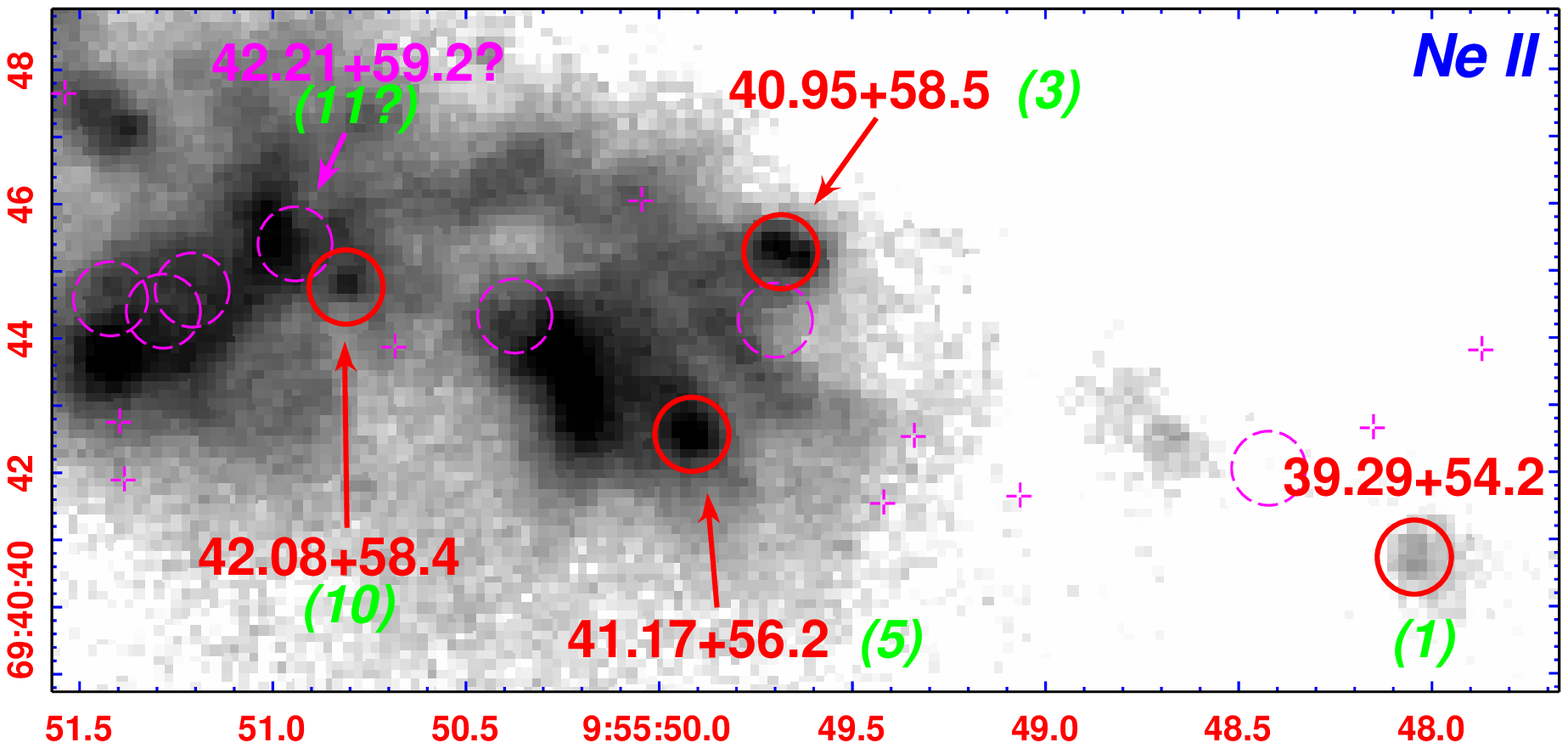}
    \includegraphics[width=8.9cm,angle=0]{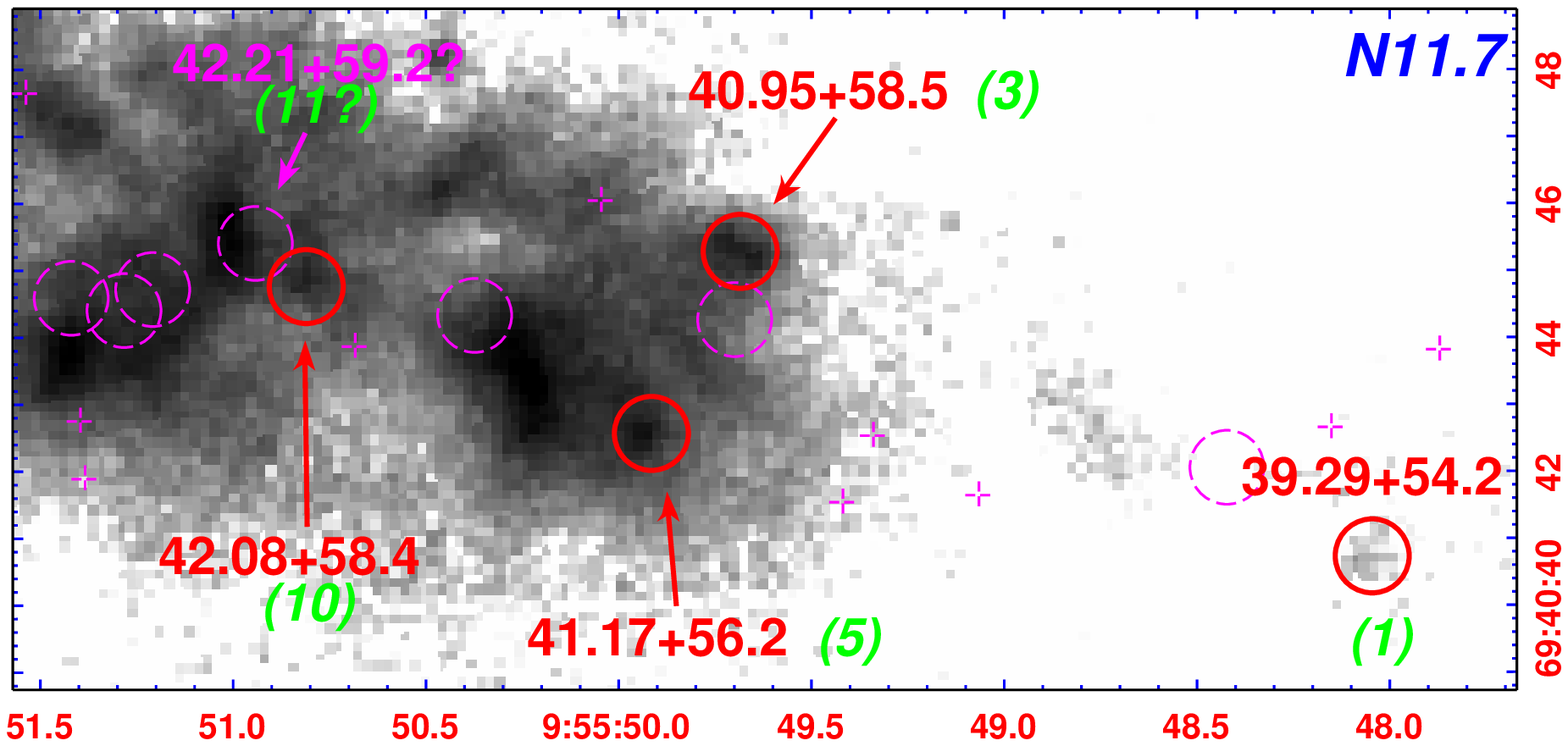}
    \includegraphics[width=6cm,angle=90]{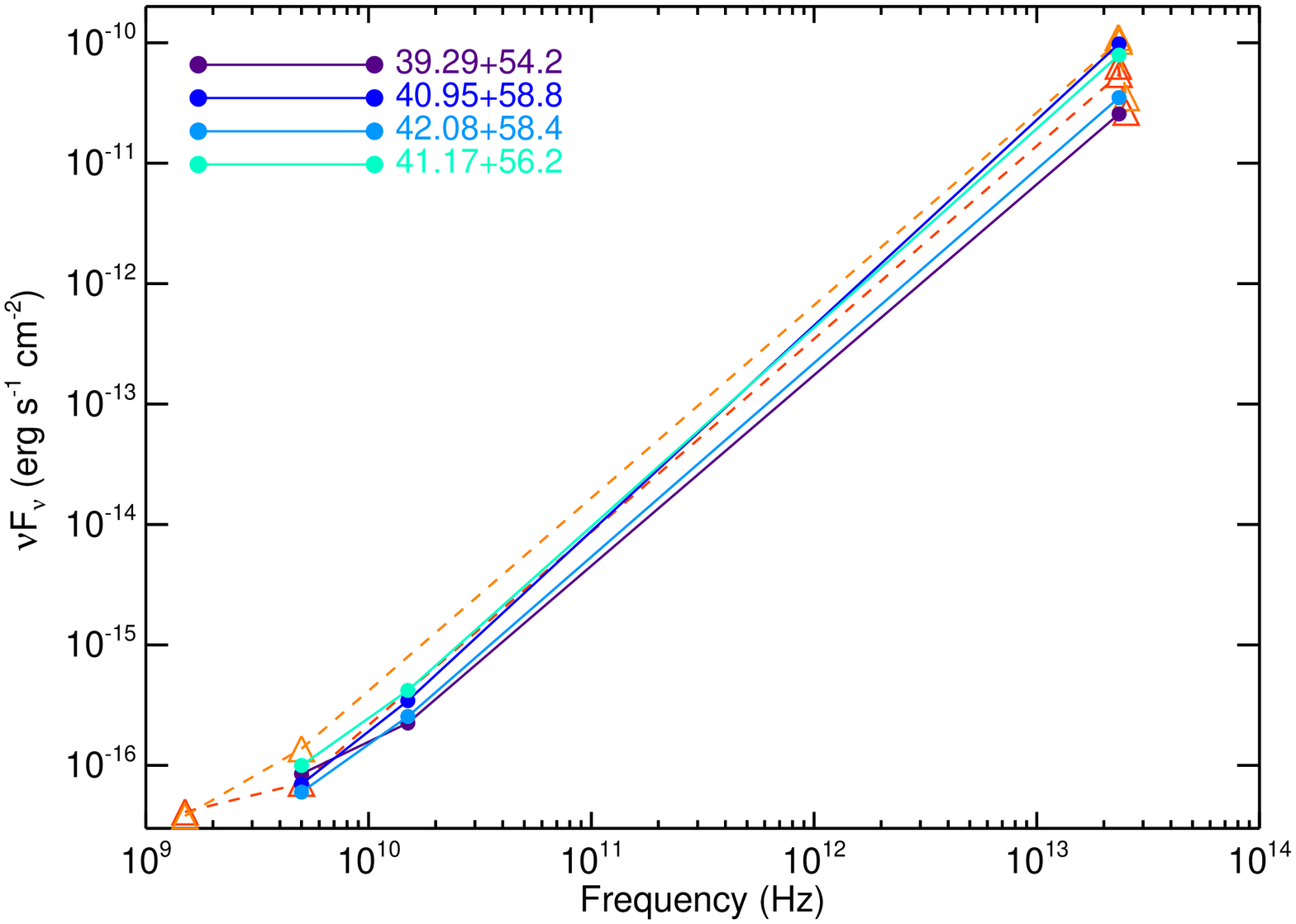}
  \end{center}
  \caption{NeII (left) and N117 (right) filter zoom-ins of radio \hii\ regions with assigned IR counterparts listed in Table~\ref{tab:compact}. The four cross-identified sources are labelled in red, and their radio positions are marked with 1\arcsec\ diameter circles. The corresponding mid-IR counterparts are labelled with brackets in green (see Table~\ref{tab:compact} and Fig.~\ref{fig:compact}. A fifth possible association of the radio source 42.21+59.2 is with mid-IR source \# 11; see text in \S~\ref{sec:image_registration}. (Bottom) Radio to mid-IR SEDs of the four cross-identified sources are shown. The dashed red and orange lines shows the SEDs for massive clusters M5 and M6 from \citet{galliano08}.}\label{fig:hiiregions}
\end{figure*}

Registration of images at multiple wavelengths (to better than half-arcsec accuracy) is a non-trivial task, given the small field-of-view of COMICS, the extended nature of the bulk of the emission and the strong dust extinction towards the core. Extensive high resolution radio studies over the years have accurately mapped many supernova remnants (SNRs) and \hii\ regions, which arguably provide the best reference system for cross-registration. We used the VLA/MERLIN 5/15 GHz study of \citet[][ hereafter M02]{mcdonald02} to search for source matches with our mid-IR NeII image. A possible solution was identified in which four radio \hii\ regions coincided with mid-IR NeII detections to within 0$\farcs$5. These four are listed in Table~\ref{tab:compact} and are highlighted in Fig.~\ref{fig:hiiregions}. The rms error of the astrometric solution fit was 0.27 pixels, or 0\farcs 036, about 10 times smaller than a single resolution element. Adopting this solution required a shift of about 0.5 arcsec from the raw Subaru astrometry baseline. The astrometry for the \n117filter\ image was tied to that determined for the NeII filter, and all four \hii\ regions are also detected in the \n117filter\ image. 

In order to check whether these mid-IR detections are consistent with being dust emission from \hii\ regions, we plot their radio to mid-IR spectral energy distributions (SEDs) in Fig.~\ref{fig:hiiregions}, and compare to the SEDs of the massive embedded star clusters identified by \citet{galliano08} in NGC~1365. The overall match is excellent given the long \lq lever arm\rq\ between the radio and infrared, and given that some natural variation amongst the strengths of the \neii\l12.81\micron\ line dominating the \neiifilter\ filter is expected. 

\begin{table*}
\begin{center}
 \begin{tabular}{lccr@{$\pm$}lr@{$\pm$}lr}
  \hline
  \#  & \multicolumn{2}{c}{Mid-IR designation}  &  \multicolumn{2}{c}{$F_{\rm NeII}$}  &  \multicolumn{2}{c}{$F_{\rm N117}$} & Cross-ID              \\ 
    &  J2000     &    B1950     &  \multicolumn{2}{c}{mJy}  &  \multicolumn{2}{c}{mJy} & and comments \\ 
  \hline
 1 & I48.05+40.5$^\dag$ & I39.30+54.0 & 110 & 12 & 19 & 3 & 39.29+54.2 (M02)\\
 2 & I48.69+42.3 & I39.94+55.9 & 95 & 12 & \multicolumn{2}{c}{--} & -- \\
 3 & I49.65+45.2$^\dag$ & I40.91+58.8 & 419 & 44 & 80 & 8 & 40.95+58.8 (M02)\\
 4 & I49.68+43.0 & I40.93+56.6 & 101 & 11 & \multicolumn{2}{c}{--} & -- \\
 5 & I49.92+42.4$^\dag$ & I41.18+56.0 & 338 & 36 & 85 & 9 & 41.17+56.2 (M02); W2 faint (AL95)\\
 6 & I50.24+43.1 & I41.51+56.7 & 771 & 81 & 233 & 24 & W2 bright (AL95) \\
 7 & I50.42+46.2 & I41.69+59.8 & 212 & 23 & \multicolumn{2}{c}{--} & -- \\
 8 & I50.44+48.2 & I41.70+61.8 & 38 & 4 & 27 & 3 & weak \neiifilter:\n117filter \\
 9 & I50.75+47.3 & I42.01+60.9 & 72 & 8 & \multicolumn{2}{c}{--} & -- \\
 10 & I50.80+44.9$^\dag$ & I42.06+58.6 & 150 & 16 & 64 & 7 & 42.08+58.4 (M02)\\
 11 & I51.01+45.7 & I42.28+59.4 & 233 & 25 & 129 & 13 & W1 faint (AL95); tentative association with 42.21+59.2 (M02)\\
 12 & I51.44+47.3 & I42.71+61.0 & 321 & 34 & 82 & 9 & on \lq bridge\rq \\
 13 & I51.44+43.7 & I42.72+57.3 & 707 & 75 & 153 & 16 & W1 bright (AL95) \\
 14 & I51.53+49.1 & I42.80+62.7 &  69 & 8 & \multicolumn{2}{c}{--} & on \lq bridge\rq\\
 15 & I52.11+49.4 & I43.38+63.1 & 559 & 59 & 145 & 15 & another component of W1 bright (AL95) \\
 16 & I52.24+47.5 & I43.52+61.2 &  55 & 8 & \multicolumn{2}{c}{--} & -- \\
 17 & I52.50+49.2 & I43.78+62.9 & 465 & 49 & 75 & 8 & western limb of \lq bubble\rq\ (E1 of AL95)\\
 18 & I52.70+45.9 & I43.99+59.6 &  42 & 6 & \multicolumn{2}{c}{--} & 44.01+59.6 (M02); AGN candidate\\
 19 & I52.80+49.3 & I44.08+63.0 & 29 & 4 & 36 & 4 & weak \neiifilter:\n117filter \\
 20 & I53.01+47.7 & I44.29+61.4 & 493 & 52 & 123 & 13 & southern limb of \lq bubble\rq\ (E1 of AL95))\\
 21 & I53.44+49.6 & I44.73+63.3 & \multicolumn{2}{c}{--} & 31 & 3 & weak \neiifilter:\n117filter \\
 22 & I54.28+54.2 & I45.57+67.9 & 142 & 16 & \multicolumn{2}{c}{--} & -- \\
  \hline
 \end{tabular}
 \caption{Parameters of detected discrete sources. Source designation is relative to (09h55m, +69d40m) and (09h51m, +69d54m) in J2000 and B1950, respectively. The prefix \lq I\rq\ in the designation refers to the fact that these are Infrared coordinates. The final column states cross-identification with the radio catalog of M02, or with diffuse structures in AL95. In some cases, other comments are given. $^\dag$These sources were used for astrometric calibration. 
\label{tab:compact}}
\end{center}
\end{table*}

Such a registration procedure is not foolproof, so we have carried out cross-checks of our astrometry against published (comparatively) low spatial resolution data mentioned previously: the \neii\ line map of AL95 and 12.4~\micron\ imaging from TG92. A near-perfect astrometric match is found with the former at the center of our field of view, but with an increasing radial offset towards the edges. The 12.4~\micron\ images of the latter show a small systematic offset to larger right ascension with respect to both our astrometry and to the line map of AL95. All of the above relative offsets lie below $\approx$1\farcs 4 and are unlikely to be significant, because the absolute position accuracies or spatial resolution available to these authors were $\sim$1\farcs 5. The above comparisons are self-consistent with our astrometric solution, and we estimate the absolute positional accuracy of our images to be better than 1\arcsec. Further detailed checks must await high resolution data and more cross-identifications over a larger field of view. 

\section{Results}

\subsection{Comparison with near-IR images}

We start with a qualitative comparison of the overall distribution of stars with that of dust. For the stellar distribution, we used a 1.6~\micron\ NICMOS image obtained from the {\em Hubble Space Telescope} archive, and its astrometry was calibrated by using the association found by \citet[][ see their Fig. 8]{kong07} for their point source designated as J095551.0 in \c\ X-ray imaging. Fig.~\ref{fig:rgb_separate} presents our \neiifilter\ image and the NICMOS 1.6~\micron\ image. These have been overlaid on to a single image in Fig.~\ref{fig:rgb}. There is a distinct anti-correlation between the visibility of stars at near-IR wavelengths, and the appearance of mid-IR dust. In particular, the group of well-known super-star clusters MGG-7, 9 and 11 (cf. Fig. 8 of \citealt{kong07}) is neatly nestled within the mid-IR gaps. Other clusters, e.g. MGG-3, 6, 8, q and other massive agglomeration of stars \citep{mccrady03}, all appear where the mid-IR emission is weakest. Heavy and patchy dust extinction is known to affect the core of the galaxy -- \av$>$25 mags \citep{willner77, rieke80, oconnell95} -- which may easily be higher within individual clumps. Using the standard interstellar extinction law \citep{riekelebofsky85}, the corresponding $H$ band extinction is expected to be $A_{\rm H}$$>$4 mags. Thus the effect of dust is sufficient to cause the anti-correlation between mid-IR and near-IR structures.

\subsection{Comparison with X-ray images}
\label{sec:chandra}
\begin{figure}
  \begin{center}
    \includegraphics[width=8.5cm]{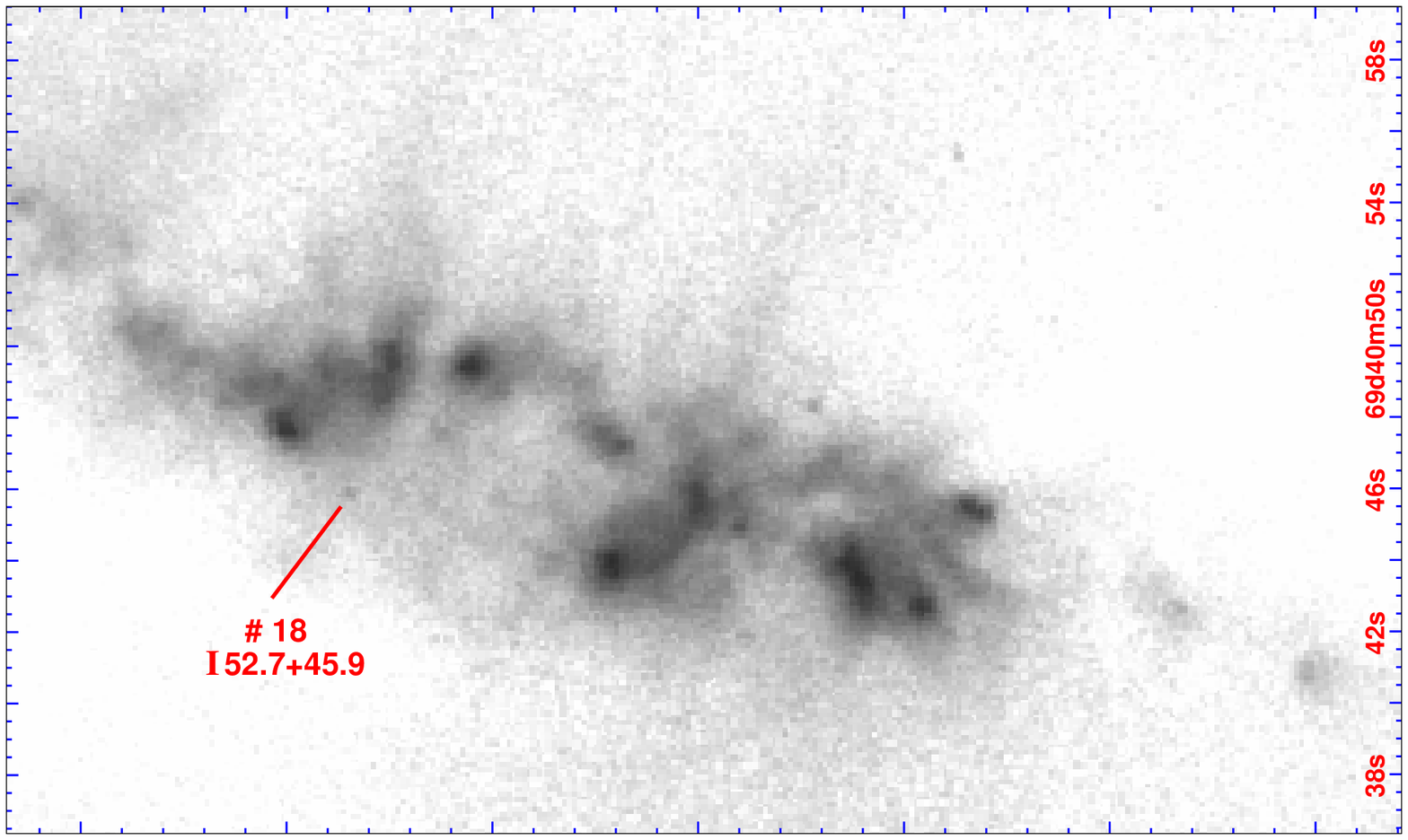}
    \includegraphics[width=8.5cm]{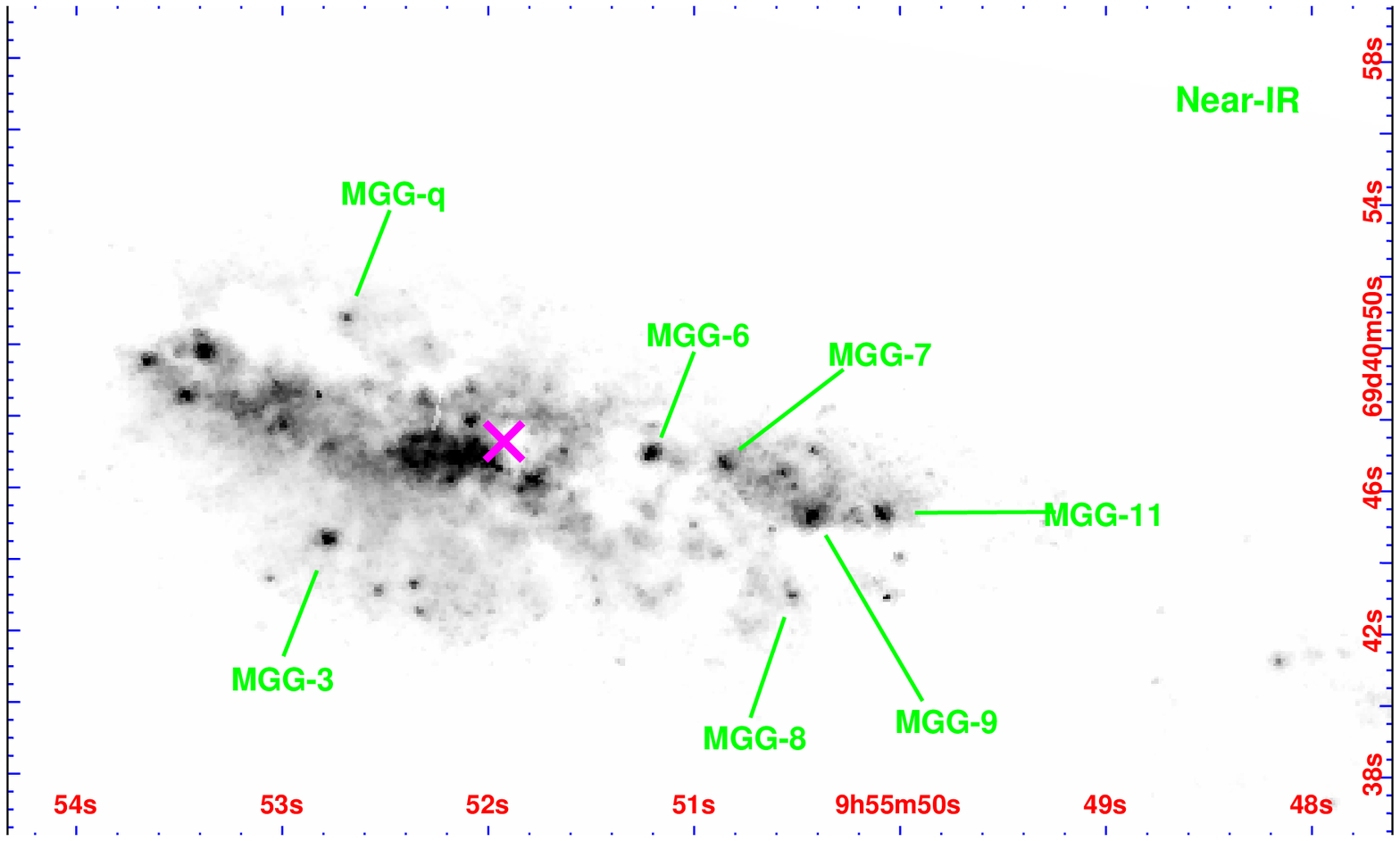}
    \includegraphics[width=8.5cm]{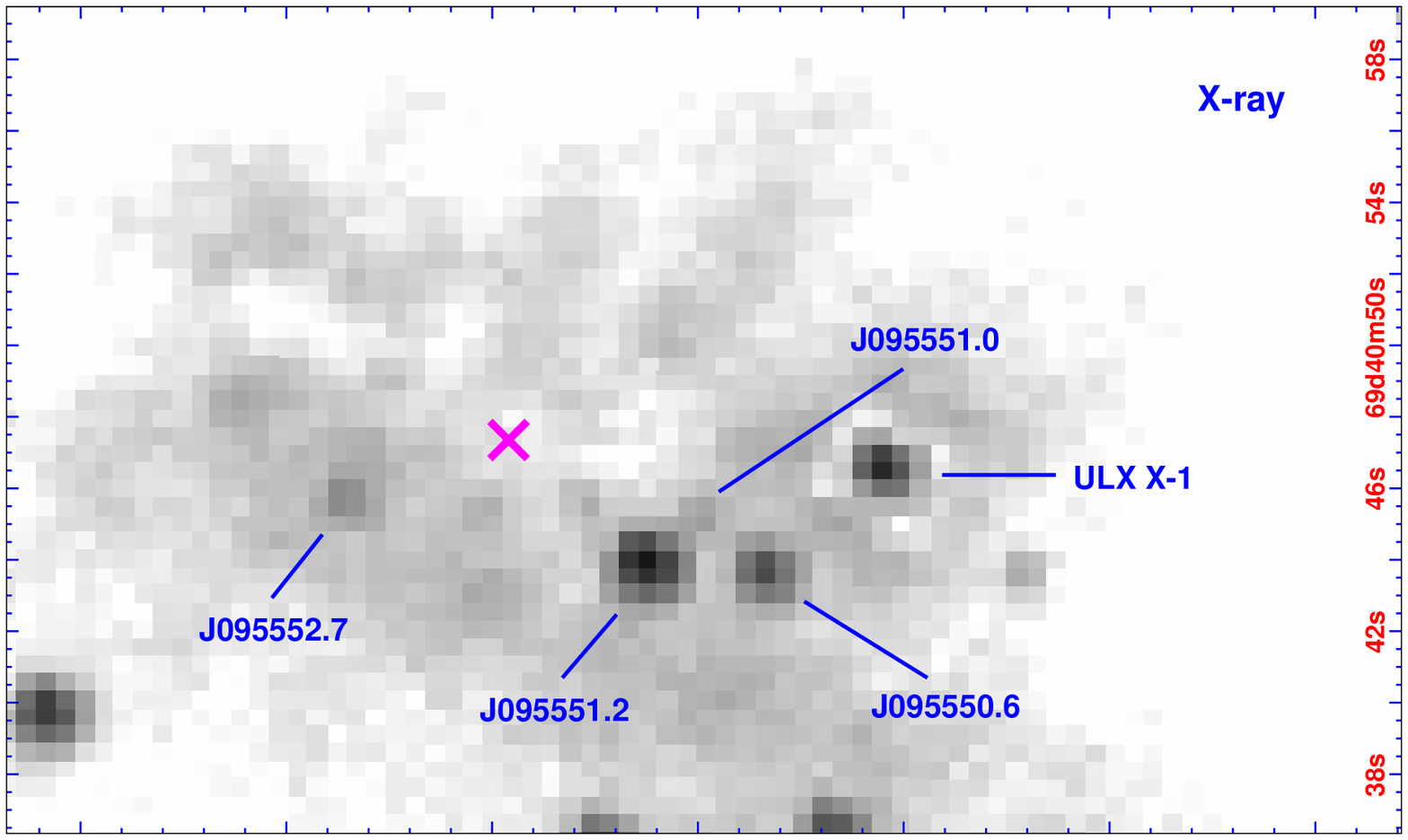}
  \end{center}
  \caption{Subaru, \hst\ and \c\ images of the core of M82, aligned to match coordinates. The cross is the radio kinematic center determined by \citet{weliachew84}. North is up and East to the left. Each image has a field of view of 35\arcsec$\times$23\arcsec. The annular depression of counts visible around the brightest \c\ source to the West is an artifact (see ~\ref{sec:chandra}).
}\label{fig:rgb_separate}
\end{figure}

\begin{figure*}
    \includegraphics[width=18cm]{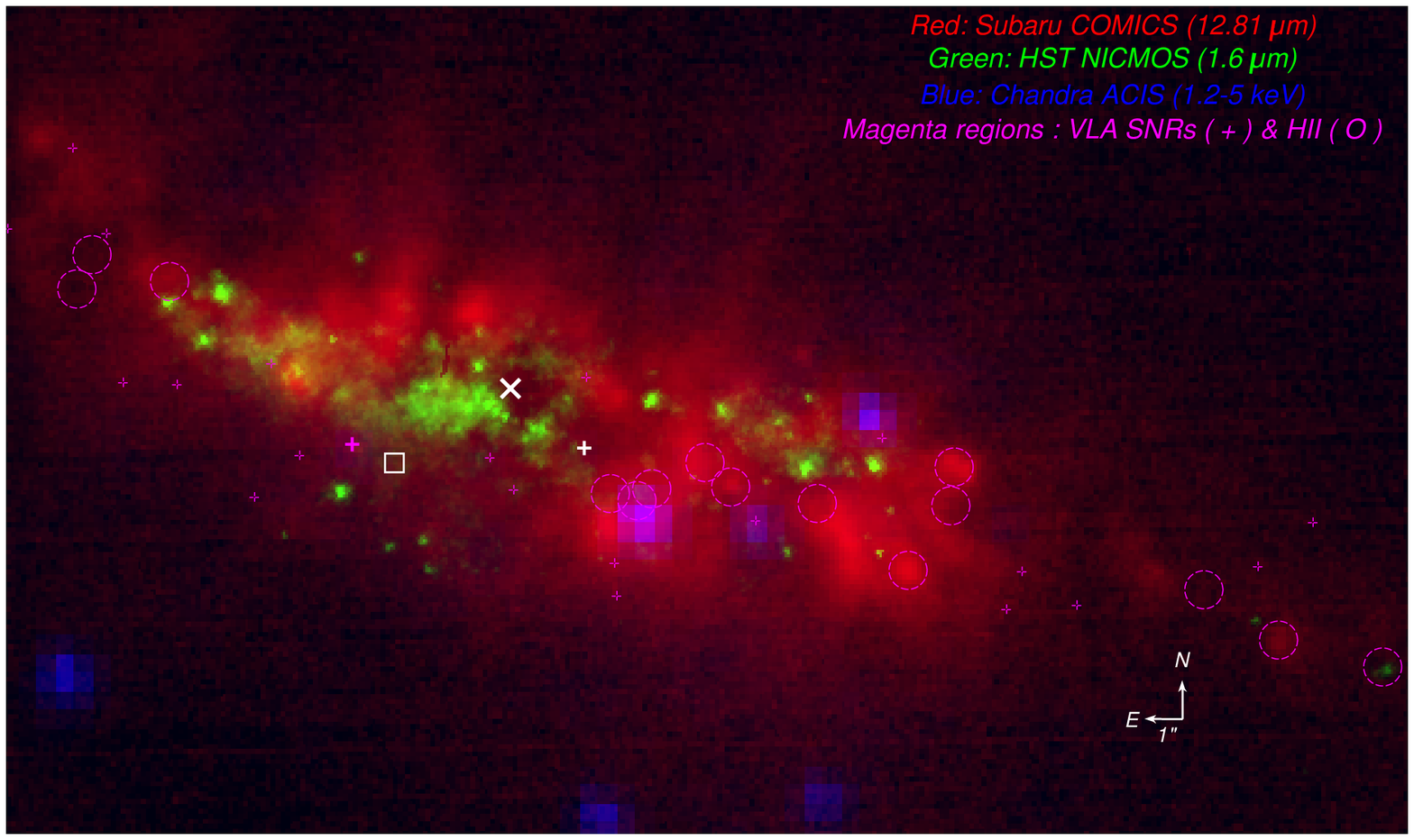}
  \caption{RGB overlay of the three panels from Fig.~\ref{fig:rgb_separate}, with radio regions plotted as dashed circles (\hii\ regions) and SNRs (plus signs). The white {\sc x} sign is the kinematic center \citep{weliachew84}. There are two heavy plus signs: about 4\arcsec\ to the East of the center, and 1\arcsec\ South is the AGN candidate in magenta (see \S~\ref{sec:nucleus}), and about 2\arcsec\ to the West and 1\arcsec\ South is SN~2008iz in white \citep{brunthaler09}. The white box marks the position of the unusual radio transient reported by \citet{muxlow10}. 
}\label{fig:rgb}
\end{figure*}

We also carried out a comparison against two long X-ray images obtained from the \c\ archive: sequences 600735 and 600736 with observation dates of 2009 Jun 24 and 2009 Jul 1, respectively. In each case the center of M82 lies on the S3 chip of Advanced CCD Imaging Spectrometer (ACIS). Results presented here use \ciao\ v4.2 and the \caldb\ v4.3.0 calibration database. The data were re-calibrated using {\sc vfaint} cleaning, with random pixelization removed and bad pixels masked, following the software \lq threads\rq\ from the \c\ X-ray Center (CXC)\footnote{http://cxc.harvard.edu/ciao} to make new level 2 events files. Only events with the recommended grades of 0,2,3,4,6 were used. The observations were free from large background flares, and after removal of time intervals when the background deviated more than 3$\sigma$ from average, the exposure times for sequences 600735 and 600736 were 118.413 and 118.054 ks, respectively. The default astrometry attached to the images was found to agree to within $\ltsim$0\farcs 3 with the data presented by \citet{kong07}, and no further refinement was done. In Figs.~\ref{fig:rgb_separate} and \ref{fig:rgb}, an exposure-corrected 1.2--5 keV merged image of the two observations is included. Note that the brightest source in the field ULX X-1 is likely to be piled up \citep[see also ][]{fengkaaret10} and a \lq halo\rq-like counts depression visible in Fig.~\ref{fig:rgb_separate} seems to be caused by this. There is also a faint readout streak extending from the source position in both directions on the ACIS data. But these effects are not relevant for our work, because the chip regions used for spectral analysis (later in \S~\ref{sec:nucleus}) do not overlap with this; nor do we use this source for astrometric calibration or any other analysis. 

There are hints of association of the mid-IR wind streamers with structures in the diffuse X-ray emission at faint levels. There is also a conspicuous dust lane of extinction which bisects the stellar distribution \citep[e.g. ][]{oconnell95}, and this appears to be associated with cold gas obscuration as well. This is manifest as a sharp decrement in the diffuse galactic X-ray emission (easily visible in the bottom panel of Fig.~\ref{fig:rgb_separate}) coincident with the near-IR deficit. The mid-IR emission avoids several spots along this band (in particular, near the dynamical center) which may be sites of extremely thick and cold absorbing material. Detailed analysis of these individual structures is left for future work.

With regard to the X-ray point sources, the most important coincidence with an mid-IR source is that of source \# 18, which will be discussed in detail in \S~\ref{sec:nucleus}. The well-known bright ultra-luminous X-ray sources (ULXs) known in the core of M82 do not show bright mid-IR counterparts. There is, however, one potential match. \citet{kong07} note that their X-ray source J095551.0 (which we use for the NICMOS astrometric calibration in the previous section) coincides with the radio source 42.21+59.2 [B1950] from M02. Our source \# 11 (I51.01+45.7 [J2000]) lies at a separation of 0\farcs 5 from the radio position. This offset, combined with the fact that the mid-IR counterpart is elongated (roughly along the north-south direction; Fig.~\ref{fig:compact}), makes cross-identification uncertain. But there are two noteworthy points: 1) M02 also found an extension for the radio counterpart along a similar direction (see also \citealt{fenech08}); 2) our mid-IR counterpart bridges the near-IR+X-ray position and the radio one. So we tentatively assign this source as the fifth \hii\ region identified in the mid-IR (it is marked in Fig.~\ref{fig:hiiregions}). As discussed by \citet{kong07}, the X-ray counterpart may be a ULX hosted within the star cluster detected by M02 in the radio (and also now by us in the mid-IR). 

\subsection{Discrete sources}

The general source population in the core of the galaxy consists of discrete sources whose emission is superposed over a diffuse radiation field. Of all detected sources, only one (source \# 8 [I50.44+48.2] in Table~\ref{tab:compact}) appears to be truly point-like based on the structural parameters returned by SExtractor and a FWHM of close to that expected for the diffraction limit. Note that the morphology of source \# 18 [I52.70+45.9], which we will discuss further in \S~\ref{sec:nucleus}, is uncertain due to its faintness and the fact that it is affected by low level read pattern noise. 

Most of the discrete sources appear extended at our high angular resolution of $\ltsim$0\farcs 4 (corresponding to a linear scale of $\ltsim$6--6.5 pc). This is in contrast to the optical and near-IR appearance of super star clusters detected in \hst\ NICMOS imaging, where most were found to have half light physical extents of $\sim$3.5 pc or less \citep{oconnell95, mccrady03}. Our detected dust features may then be a result of large scale outflows from the starburst regions. 

The measured source centroids and fluxes are listed in Table~\ref{tab:compact}, and the sources are identified in Fig.~\ref{fig:compact}. There is good overall agreement in the emission structures between the two images, though the NeII filter shows a greater number of significant individual detections as compared to \n117filter, especially towards the outer parts of the field of view. Many sources have NeII filter fluxes higher than the corresponding \n117filter\ filter fluxes by factors of several at least, meaning that the \neii\l12.81\micron\ dominates in these cases. This is true for all the \hii\ regions cross-matched with the radio catalog of M02, which is consistent with the fact that this emission line is a strong star-formation indicator, and also why more sources are detected in the \neiifilter\ filter.

In Table~\ref{tab:compact}, we comment on some source properties and associations with ionized gas sources identified by AL95. In particular, we have found additional structure to their \lq E1, W1 and W2\rq\ peak emission sites on the eastern and western limbs, and on the western ridge of \neii\ line emission. A couple of sources are also found on the faint \lq bridge\rq\ of emission connecting the eastern and western limbs of the ring. Finally, two of the tabulated sources (\# 19 and \# 21) have \neiifilter\ fluxes of less than 1.5 times the continuum flux in the \n117filter\ filter, much less than in other cases. This may suggest atypical SEDs with continua rising to shorter wavelengths. Upon examining Fig.~\ref{fig:compact}, though, it seems that extended emission is present in the \neiifilter\ filter at the positions of both these sources, but SExtractor is simply unable to isolate it from the highly clumpy diffuse background. In fact, simulating the addition of a weak point source at the position of source 21 in the \neiifilter\ filter suggests a limit of $\sim$90 mJy which, in turn, implies a NeII/N11.7 flux ratio consistent with the distribution of flux ratios for detected sources in Table~\ref{tab:compact}.

\subsection{Diffuse features and galactic emission}
\label{sec:diffuse}

Underlying the discrete sources is a diffuse emission field from the galaxy which completely dominates the radiated flux. Integrating all observed photon counts over the COMICS field of view yields total fluxes of 55 and 108 Jy in the \n117filter\ and NeII filters, respectively. Systematic uncertainties from detector cosmetics and the standard star absolute calibration dominate here (\S~\ref{sec:photometry}), for which we allow for 10\%\ variations along the full field. The 11.7~\micron\ flux is within a factor of 2 of the continuum flux reported by \citet{beirao08} for their \lq total\rq\ aperture. This comparison is likely to be full consistent, given the difference in aperture sizes and positions used, perhaps some contribution of the 11.2~\micron\ polycyclic aromatic hydrocarbon feature to our broad-band photometry, and absolute cross-calibration uncertainties. Removing the combined flux of the discrete sources identified in Table~\ref{tab:compact} gives 53 and 106 Jy for the two filters. This diffuse emission possesses substantial sub-structure, of which we highlight a few specific aspects. 

\begin{figure}
  \begin{center}
    \includegraphics[width=8.5cm]{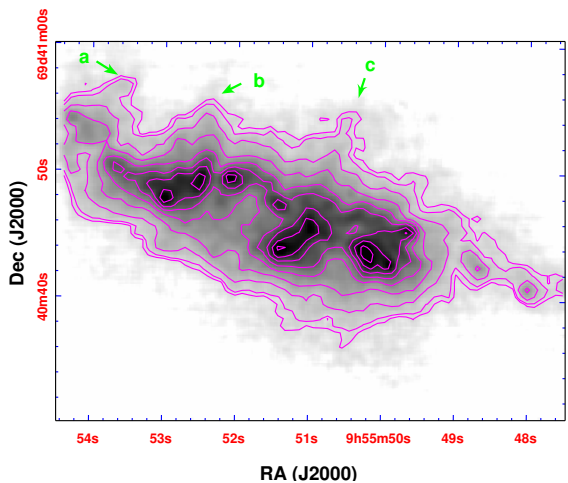}
  \end{center}
  \caption{\neiifilter\ filter image smoothed with a convolution filter in order to highlight diffuse features. Some of the main features are labelled: \lq a\rq\ and \lq b\rq\ resemble the projected limbs of a bubble, and \lq c\rq\ is a long chimney-like structure. The contours are plotted in a \lq square-root\rq\ scaling in order to highlight faint features, with 10 levels ranging from 1.3 to 24 mJy pixel$^{-1}$.}\label{fig:basicimages_smooth}
\end{figure}

The most striking features of our high resolution imaging are several wind \lq streamers\rq, which stand out better in Fig.~\ref{fig:basicimages_smooth} which has been smoothed by convolution with a square top hat kernel 5$\times$5 pixels (0$\farcs$67$\times$0$\farcs$67) wide, in order to enhance faint diffuse features. Most of these are visible on the northern side, with an orientation roughly perpendicular to the major axis of the galaxy. The longest of these in our image is a chimney-like straight structure, with a linear extent of over 7\arcsec, or about 120 pc. Chopping in the mid-IR cuts out a lot of extended emission. So, we are undoubtedly seeing only the inner regions of much larger streamers and chimneys which have been detected on wider scales \citep[e.g. ][]{engelbracht06, kaneda10}.

Towards the north-east of the field of view, two structures extend outwards and curve in towards one another, resembling the edges of a limb-brightened bubble. Assuming a circular shape, the projected diameter of this feature is $\approx$8$\arcsec$, corresponding to a physical size of $\approx$140 pc. Limb-brightening is much less pronounced near the top, which may suggest wind rupture, if real. Ruptured bubbles are a natural consequence of over-pressurization of outflowing hot gas in the superwind model which explains the burst of star formation in the core of M82 \citep[e.g. ][]{chevalierclegg85, heckman90}. If there is entrained clumpy dust within this outflow, the dust distribution will follow that of the gas. On the other hand, there is no obvious evidence from multi-wavelength data that this bubble is filled in with hot plasma. Its reality as a bubble or as two separate streamers need confirmation from further observations. 

These various features do not point back to a single site of energy ejection. The starburst in M82 is known to occur in a \lq ring\rq, rather than a compact nuclear concentration \citep[e.g. ][AL95]{nakai87}, and the mid-IR emitting hot dust indeed traces out portions of this ring. Multiple burst events of dust expulsion along the entire inner $\sim$500 pc region of this ring are required to explain the data. 

\subsection{Dust temperature and mass}

The mass of the hot, mid-IR emitting matter may be estimated by assuming optically-thin thermal radiation from uniform spherical dust grains. The largest uncertainty in this estimate is the unknown and spatially-variant dust temperature. \citet[][hereafter TH80]{telesco80} were able to fit a modified black body emission model with a temperature of $\sim$45 K for the dust emitting at far-IR wavelengths of 41--141 \micron\ over the central $\sim$900 pc region. They also found the observed fluxes at shorter wavelengths from 2--21~\micron\ to overestimate the thermal model predictions by many orders of magnitude, meaning that hotter dust components characterize the mid and near-IR emission. 

\begin{figure}
  \begin{center}
    \includegraphics[width=8.5cm]{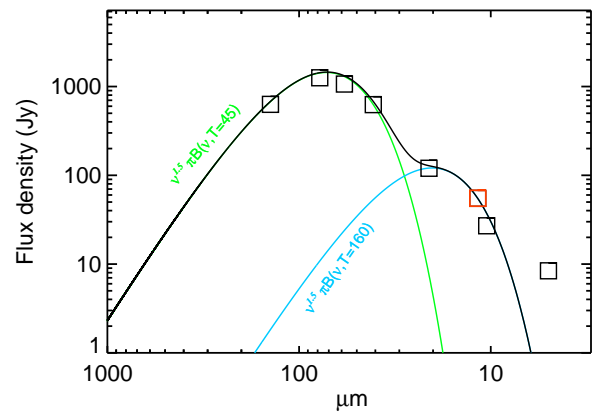}
  \end{center}
  \caption{Archival wide beam fluxes of the integrated emission from the central regions of M82 from \citet{telesco80} (beyond 40~\micron) and \citet{riekelow72} (5, 10.5 and 21~\micron) overplotted with our integrated COMICS 11.7~\micron\ flux (red box). The green and blue curves represent modified black bodies with temperatures of 45 K (cf. Fig.~1 of TH80) and 160 K, respectively. The black curve is the sum of these two.}\label{fig:dustbb}
\end{figure}

In Fig.~\ref{fig:dustbb}, we overplot the 5--41~\micron\ fluxes and 45~K thermal model from TH80, along with our integrated \n117filter\ filter flux. There is some non-negligible uncertainty resulting from varying beam sizes and centroids amongst the data plotted in this figure so we do not carry out a detailed fit. But a qualitative comparison is reasonable if the flux is mostly centrally-concentrated. If a two-temperature-only model is assumed, we find that a modified black body with flux density varying as $\nu^{1.5}B_{\nu,T}$ (where $B_{\nu,T}$ is the Planck function at frequency $\nu$ and temperature $T$) is a fair representation of the 10.5--21~\micron\ integrated source fluxes for $T$$\sim$160 K. The emissivity power-law index of 1.5 has been assumed to be identical to that found for the cooler dust by TH80. This model is also plotted in Fig.~\ref{fig:dustbb} (normalized to our 11.7~\micron\ data), and shows that the 5~\micron\ excess requires other even hotter dust components, if this emission is thermal in origin.

Using this model, the dust mass may be estimated by employing the infrared emissivity relation of \citet{hildebrand83},

\begin{equation}
M_d=\frac{4\pi a \rho D^2}{3}\frac{F_\nu}{Q_{\nu,a}\ \pi B_{\nu, T}}
\end{equation}

\noindent
where $M_d$, $D$, $a$ and $\rho$ are the dust mass, galaxy distance, grain radius and specific density, respectively. $F_\nu$ and $Q_{\nu,a}$ are the observed flux density and the grain emissivity. Assuming emission from the grain surface into all hemispherical solid angles gives the factor of $\pi$. Taking canonical values of the size and density for a graphite grain composition gives 

\begin{eqnarray}
 M_d &=& 1.44\times 10^{-27} \left( \frac{a}{0.1\ \mu{\rm m}}\right) \left( \frac{\rho}{2.26\ {\rm g\ cm}^{-3}}\right) \nonumber \\
   &&\times  \left( \frac{D}{{\rm pc}}\right)^2  \left( \frac{F_\nu}{\rm 1~Jy}\right) Q_{\nu, a}^{-1}\ B_{\nu, T}^{-1}\ \  {\rm M}_{\odot} 
\end{eqnarray}

\noindent
for $B_{\nu, T}$ expressed in W m$^{-2}$ Hz$^{-1}$ sr$^{-1}$. A mean 11.7~\micron\ emissivity of $Q_{\nu,a}$$\approx$0.009 for graphite grains with diameter of $a$=0.1~\micron\ is used \citep{draine85}. The observed total flux and model temperature then imply $M_d$$\sim$950~\Msun\ for the hot dust. In comparison with the mass of cooler 45 K dust determined by TH80, this is about 500 times lower. 

\section{Candidate AGN}
\label{sec:nucleus}

Our target \# 18 (I52.70+45.9 [J2000]) has a uniquely interesting story. In our astrometry-corrected images, it is found to be coincident (within $\approx$0\farcs 1) with the radio source 44.01+59.6 [B1950] from \citet[][ and references therein]{mcdonald01}. Based upon several lines of evidence, it has been suggested that this may be a weak active galactic nucleus (AGN) in M82. The evidence includes the atypical radio SED, detection of OH maser satellite lines and also a possible elongated radio jet (\citealt{wills97, seaquist97, wills99}). Other detailed studies have shown that its SED is not atypical for a SNR, and it has an expanding shell with an expansion velocity of 2700\p400 km s$^{-1}$ \citep{allenkronberg98, fenech08}. Several hard X-ray AGN searches have been carried out over the years \citep{tsuru97, ptakgriffiths99}, and a potential counterpart has been detected with \rosat\ \citep[][ cf. their source X-3]{stevens99} and even \einstein\ \citep[][ who also suggested this as a possible nucleus, based on its proximity to the 2.2~\micron\ core]{watson84}. But there is considerable diffuse soft X-ray emission in this region, and the high resolution of \c\ is required for resolving the complex unambiguously \citep[e.g. ][]{matsumoto01}. No unambiguous and strong X-ray:radio association to our knowledge has been reported to date. The source lies close to, but not exactly at the dynamical center of the galaxy. It lies 4\farcs 1 (70 pc) and 2\farcs 1 (36 pc) from the radio and optical kinematic centers quoted by \citet[][ see also \citealt{oconnell78}; errors are $\approx$1\farcs 5--2\arcsec\ in each coordinate]{weliachew84}. In summary, the nature of this object still remains uncertain. 

In Fig.~\ref{fig:nucleus}, we present zoom-in images of the source position in the mid-IR (our NeII filter), the near-IR (NICMOS 1.6~\micron), and in X-rays (\c\ ACIS 0.5--2 keV and 3--7 keV). There is no point-like source in the near-IR, but it is detected significantly in our NeII image. A source also clearly appears in X-rays, especially at hard energies. We designate this source as CXOU J095552.7+694046 (with J2000 coordinates of 09:55:52.7+69:40:45.8), or J095552.7 for short. Its position relative to other X-ray sources within the core of M82 discussed by \citet{kong07} is shown in the bottom panel of Fig.~\ref{fig:rgb_separate}. We now present analysis of the new long-exposure \c\ archival data analyzed herein. In the next section, we will also examine the broad-band SED and comment on the source nature. 

\subsection{Chandra X-ray data} 
\label{sec:xspec}

First, we extracted a radially-averaged surface brightness profile of the source, and compared this to a simulated PSF image. The PSF was constructed using the \c\ ray tracing package {\sc chart}\footnote{http://cxc.harvard.edu/chart} assuming observational parameters including nominal and source coordinates, roll angle, and ACIS-S chip identical to those for the analyzed data (the sequence 600735 was used for this comparison). The ray trace was then projected onto the detector plane and a PSF events file generated using {\sc marx}. The profile of the source for photons extracted over the hard 3--7 keV energy range, compared to the PSF expected for 4.5 keV photons, is shown in Fig.~\ref{fig:agn}. A constant background level corresponding to the outermost radial bin has been removed. The source appears to be consistent with a point source, at least within the core. There may be evidence of an extended \lq wing\rq\ around 1\arcsec\ radius, but systematic effects of the (position-dependent) underlying galactic background cannot be ruled out. 

\begin{figure*}
  \begin{center}
    \frame{\includegraphics[width=6.5cm]{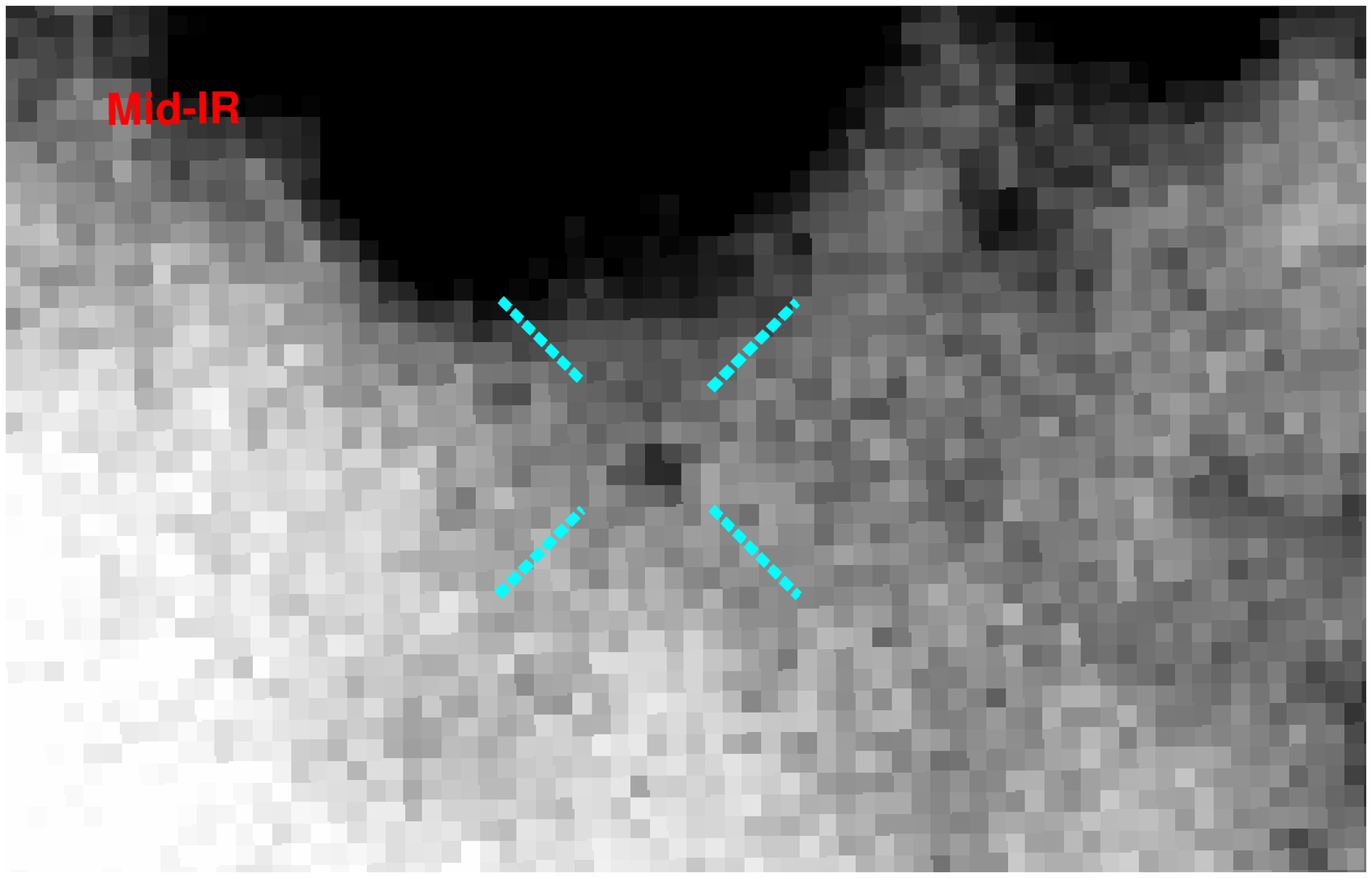}}
    \frame{\includegraphics[width=6.5cm]{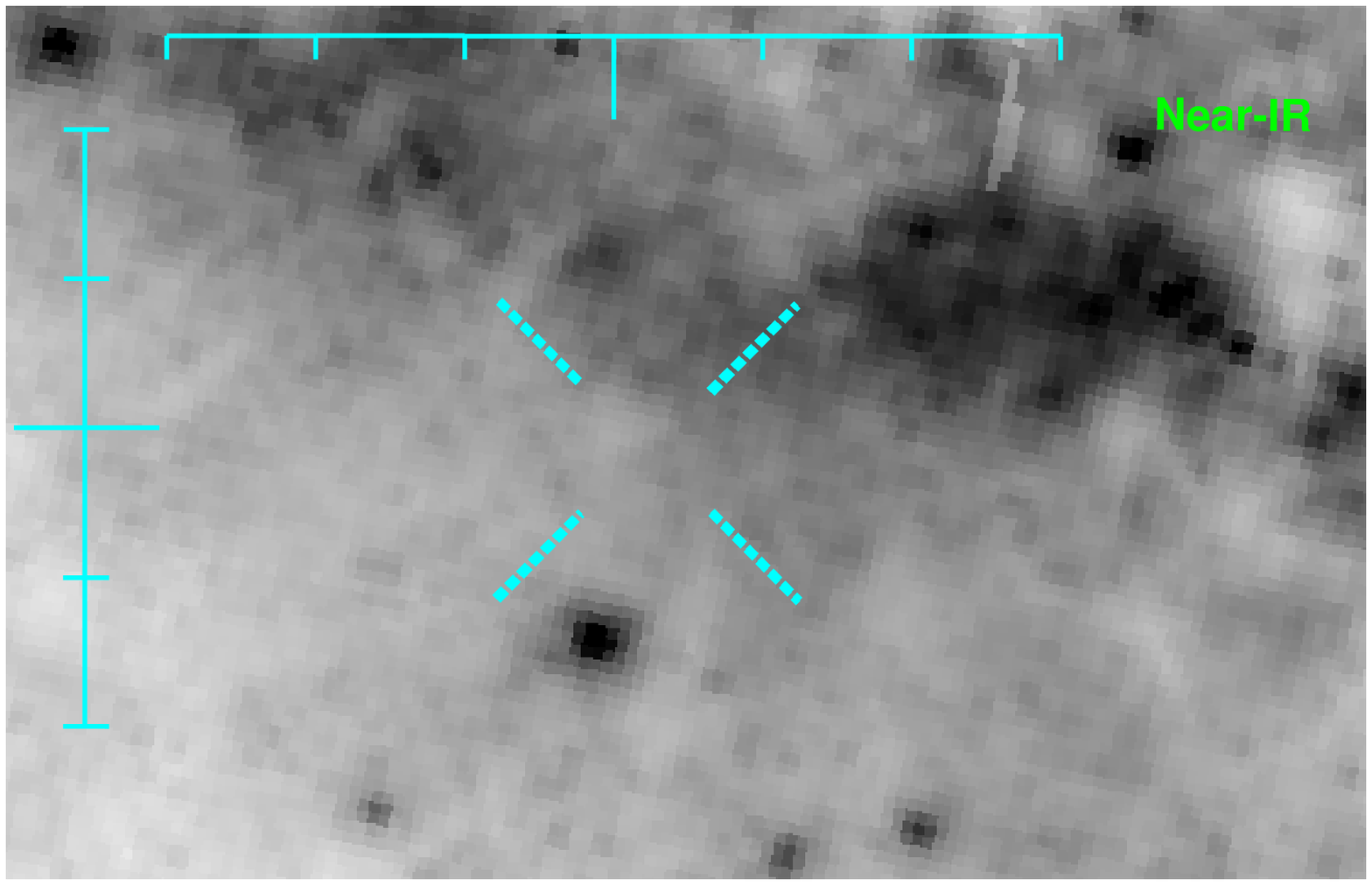}}
    \frame{\includegraphics[width=6.5cm]{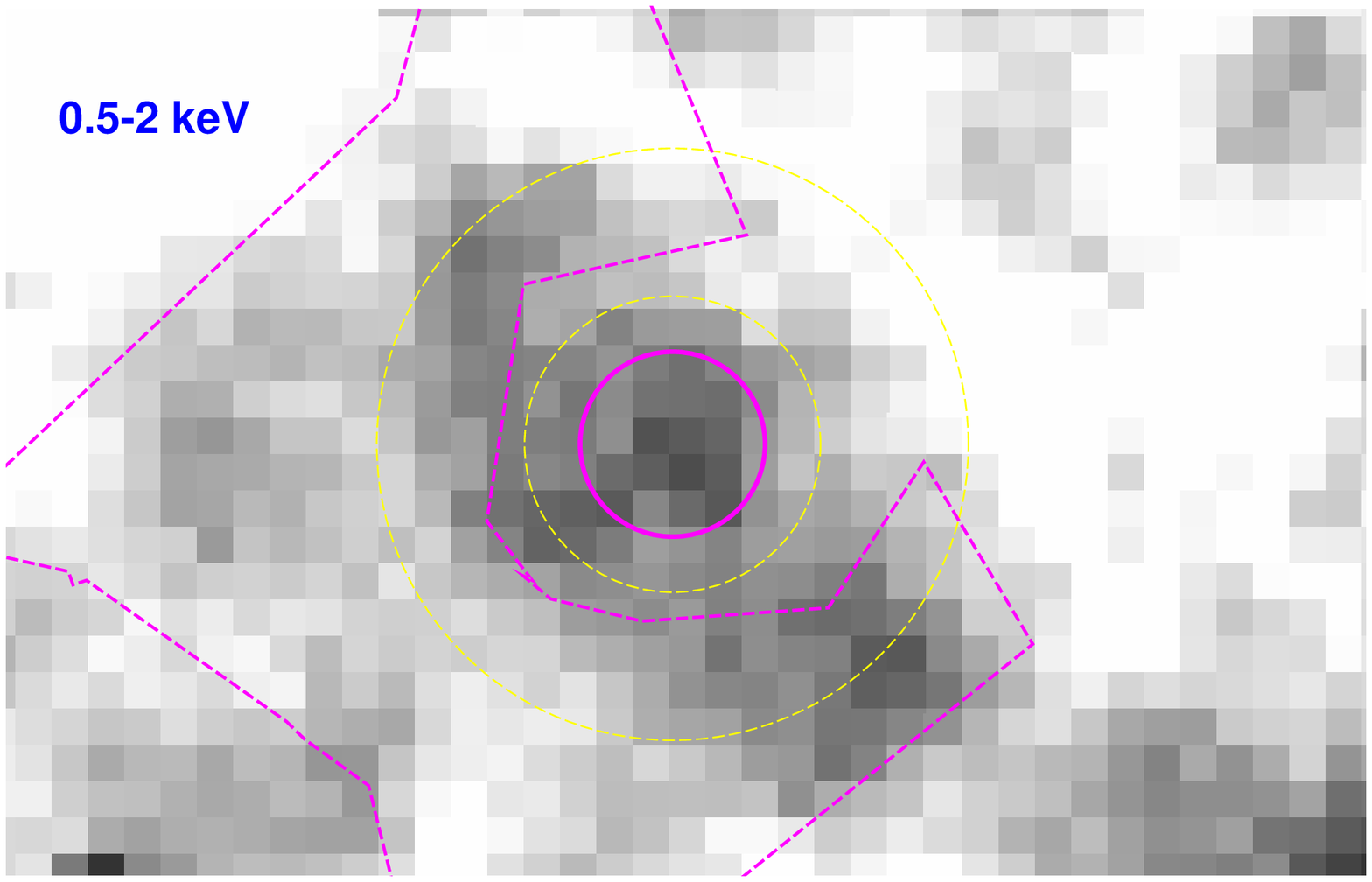}}
    \frame{\includegraphics[width=6.5cm]{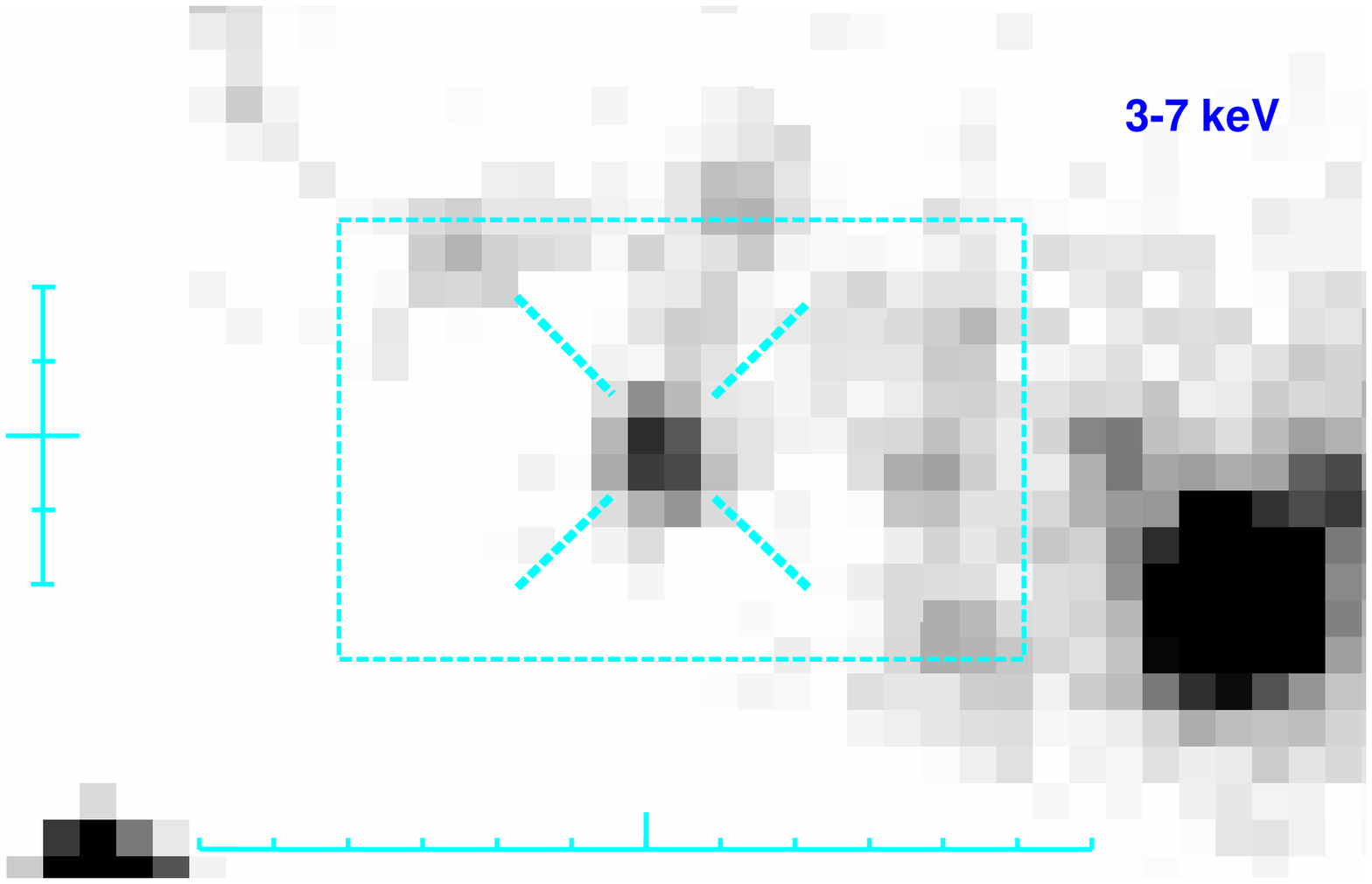}}
  \end{center}
  \caption{Magnified images (9\arcsec wide $\times$ 6\arcsec\ high) on the position of the AGN candidate (source \# 18) in the mid-IR (top left; COMICS NeII filter) and near-IR (top right; {\em HST} NICMOS  F160W filter). The X-ray images at the bottom are for energy ranges of 0.5--2 keV (left) and 3--7 keV (right), respectively. Because the PSF of ACIS is larger than in the IR, both X-ray images are zoomed out by a factor of two; the dashed box in the bottom-right panel shows the field size of the IR panels. The scale bar notches denote 1\arcsec\ offsets from the source position in both RA and Dec. The regions used for spectral extraction of the target and background are shown in the 0.5--2 keV image as the magenta circular and polygonal apertures, respectively. A secondary annular background region used for a consistency check is also shown in yellow.\label{fig:nucleus}
}
\end{figure*}

\begin{figure*}
  \begin{center}
    \includegraphics[width=8.cm,angle=0]{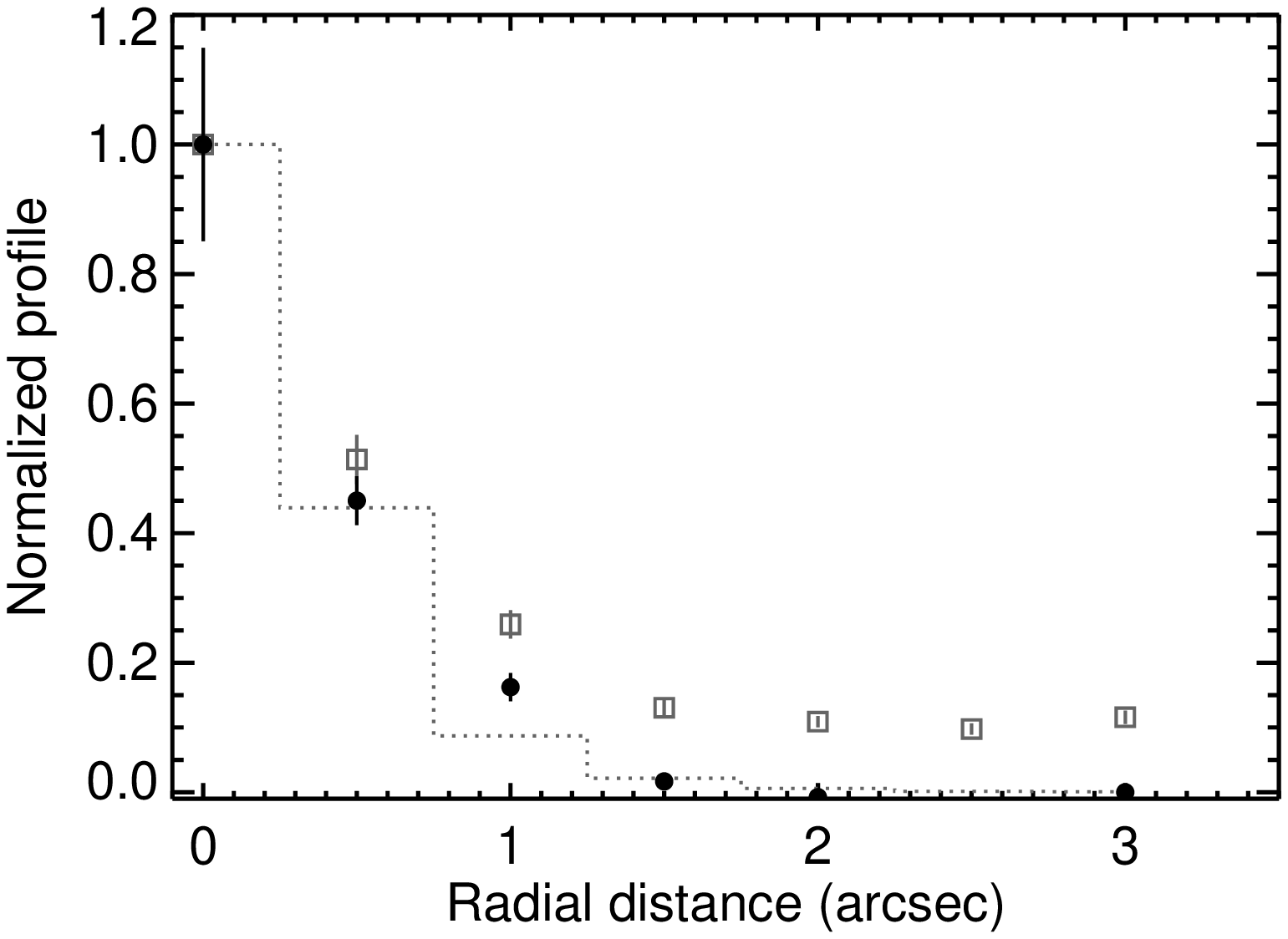}
    \includegraphics[width=7.cm,angle=270]{fig10b.ps}
    \includegraphics[width=7.cm,angle=270]{fig10c.ps}
  \end{center}
  \caption{ s
{\textbf{\textsl{(Top})}} Azimuthally-averaged \c\ ACIS-S radial surface brightness profile of J095552.7 (points with error bars). The empty gray squares show the normalized counts profile before background subtraction. The filled black circles denote the normalized net profile, and these can be compared to the expected PSF from a point source (histogram). The source profile is extracted from the 3--7 keV image, and the PSF from ray-tracing of 4.5 keV photons. 
{\textbf{\textsl{(Bottom Left})}} X-ray spectrum and residuals to a fit with two absorbed thermal plasma models. Data from both analyzed observations (black and red for sequences 600735 and 600736, respectively) are included. The two thermal models are shown as the dotted lines.
{\textbf{\textsl{(Bottom Right})}} Ungrouped spectrum around the Fe K line and best fit model from the left panel (rebinned to obtain a minimum 2$\sigma$ significance for plotting only).
\label{fig:agn} }
\end{figure*}

Next, the X-ray spectrum of the source was extracted. A circular aperture of radius 1\farcs 25 was used to accumulate source counts. The diffuse galactic flux around the source is clumpy and affected by cold gas absorption to its immediate north. An irregularly-shaped polygonal region surrounding the source was selected for background subtraction, avoiding nearby point sources as well as strong absorption to the north-west. These regions are shown in Fig.~\ref{fig:nucleus}. For spectral extraction the \ciao\ task {\tt psextract} was followed by {\tt mkacisrmf} and {\tt mkarf} to apply the latest calibrations. Models were fitted jointly to the two separate observations. Fitting was performed in \xspec\ \citep{xspec} and includes absorption along the line of sight in our Galaxy fixed at a column density of \nh= 4$\times$10$^{20}$ cm$^{-2}$ \citep[based on the data by ][]{dickeylongman90}\footnote{using the {\tt colden} program provided by CXC}. 

A minimum grouping of 30 counts per spectral bin was applied for fitting and uncertainty determination with the $\chi^2$ statistic. A fit to the resultant 0.7--6.8 keV spectrum (ignoring \lq bad\rq\ bins and energies below 0.5 keV) with a single power law component results in a hard power law with photon index $\Gamma$=1.11\p 0.08 (all uncertainties are quoted at 90\% confidence), but with an unacceptable goodness of fit $\chi^2$=175 for 100 degrees of freedom (dof), with residuals suggesting the presence of emission features below 4 keV including those from highly ionized Si, Ne and Ar. Several models, including a combination of absorbed power laws, or a power law combined with a thermal ({\sc apec}) plasma model \citep{apec} were tried. Although some of these yield statistically-acceptable fits, the power law slope above 3 keV turns out to be very soft, with $\Gamma$ restricted to $\gtsim$2--3 in all cases. Furthermore, residuals suggesting additional emission bands still remain unaccounted for.

The best model is a combination of two absorbed thermal plasmas, with temperatures of 0.6 keV and 2.6 keV, absorbed by significantly different columns of \nh=8$\times$10$^{21}$ and 4.3$\times$10$^{22}$ cm$^{-2}$, respectively. With abundances fixed at Solar, assuming the \citet{lodders03} abundance table for both temperature components, a $\chi^2$/dof=101.5/96 is found. The spectrum with this best fit is shown in Fig.~\ref{fig:agn} and the parameters are listed in Table~\ref{tab:xspec}. Adopting only a single absorber produces an unacceptable fit. Regarding abundances, assuming a yield appropriate for SN II instead \citep[][ averaged over a Salpeter initial mass function from 10 to 50 \Msun, with a progenitor metallicity of Z=0.02]{nomoto06} produces a fit similar to the best one. Letting the abundances of the two components vary also does not improve $\chi^2$ significantly. On the other hand, those for SN Ia \citep[][ their W7 model]{iwamoto99} yield an unacceptable fit. Residuals suggest a high abundance for specific elements, but these are not strictly required by the fit with the present data quality.

The observed flux is $F_{0.5-10\ \rm keV}$=8.9(\p 0.4)$\times$10$^{-14}$ erg s$^{-1}$ cm$^{-2}$, corrected for Galactic absorption only. Isolating the low and the high temperature thermal plasma components, and correcting also for the respective fitted intrinsic obscuring columns, yields $F_{0.5-10\ \rm keV}$=1.2(\p 0.1)$\times$10$^{-13}$ erg s$^{-1}$ cm$^{-2}$ and $F_{0.5-10\ \rm keV}$=2.4(\p 0.2)$\times$10$^{-13}$ erg s$^{-1}$ cm$^{-2}$, respectively. Though these two values are of the same order of magnitude, the high temperature component completely dominates above $\sim$2 keV. Errors refer to those for 90\%\ confidence intervals determined using the {\sc cflux} convolution model in \xspec. Adding these fluxes gives a total deabsorbed flux of $F_{0.5-10\ \rm keV}$=3.6(\p 0.2)$\times$10$^{-13}$ erg s$^{-1}$ cm$^{-2}$, or $L_{0.5-10\ \rm keV}$=5.3(\p 0.3)$\times$10$^{38}$ erg s$^{-1}$. 

Lastly, we point out that the diffuse galactic emission underlying the source has a soft spectrum, resulting in some additional systematic uncertainty in the source morphology and soft band spectral parameters. The source appears to have a relatively mild peak excess in Fig.~\ref{fig:nucleus}. We carried out an additional check of the soft band spectrum by changing the background extraction regions to an annulus encompassing only background counts from the immediate source vicinity. This did not remove the need for the soft thermal component, though the confidence intervals of the fitted parameters did change, as expected. The hard band, however, is not affected significantly by the galactic background. 

\subsection{Fe K emission line}

The high temperature plasma should also emit a strong ionized Fe line around 6.7 keV, near the limit of the binned energy range available. In order to analyze this line, in Fig.~\ref{fig:agn} we show the ungrouped spectrum around 6.7 keV, overplotted with the best fit model determined above (binning is applied for display purposes only). The default model clearly produces a line feature also seen in the data. 

We estimate its significance in a couple of ways. Accumulating net source counts over 6.5--6.9 keV after subtracting off the model continuum measured at the ends of this energy range, and comparing these counts with the total (background inclusive) counts in the same range, yields a simple Poisson signal:noise ratio of 2.4 and 3.1 for the line feature in the two \c\ data sets, respectively. Another estimate can be obtained by parametrizing the continuum as a power-law locally, and then fitting the line as a Gaussian superposed on this. The energy range of 5--8 keV is used in order to estimate the continuum, and we assume a Gaussian width $\sigma$ fixed at 10 eV. The $C$-statistic is the appropriate one to use for ungrouped data \citep{xspec}. We find a line energy of 6.68(\p 0.03) keV and a 90\%\ equivalent width range of 0.6--4 keV; in other words, the line is significantly detected (despite large uncertainties on its strength) and its center energy is consistent with highly ionized Fe at 6.7 keV, and not the neutral line energy of 6.4 keV. 

\begin{table}
 \begin{center}
   Spectral fit for the X-ray counterpart to source \#18
   \begin{tabular}{lcr}
     \hline
     \multicolumn{2}{c}{Model components} & Value \\
     \hline
     {\sc phabs}     &                    & 4$\times$10$^{20}$ cm$^{-2}$\\
     &&\\
     {\sc wabs$_1$}        &                    & 7.9$_{-1.1}^{+1.0}$$\times$10$^{21}$ cm$^{-2}$\\
     &&\\
                     &       $kT_1$         & 0.58$_{-0.09}^{+0.07}$ keV\\
     {\sc apec$_1$}      &    Abundance       & 1\\
                     &       norm$_1$         & 7.42$_{-2.06}^{+2.65}$$\times$10$^{-5}$$^{\dag}$\\
     &&\\
     {\sc wabs$_2$}        &                    & 4.28$_{-1.05}^{+1.16}$$\times$10$^{22}$ cm$^{-2}$\\
     &&\\
                     &       $kT_2$         & 2.56$_{-0.48}^{+0.82}$ keV\\
     {\sc apec$_2$}      &    Abundance       & 1\\
                     &       norm$_2$         & 1.99$_{-0.66}^{+0.77}$$\times$10$^{-4}$$^{\dag}$\\
     \hline
   \end{tabular}
 \end{center}
 \caption{X-ray joint spectral fit to the two \c\ archival data sets analyzed. The model is {\sc phabs (wabs$_1$*apec$_1$ + wabs$_2$*apec$_2$)}, with abundance fixed to that of \citet{lodders03}. Errors correspond to 90\%\ parameter uncertainties on a single interesting parameter, when not fixed. This model returns a statistic of $\chi^2$/dof=101.5/96. $^\dag$The units of norm are 10$^{14}$ cm$^{-5}$. 
  \label{tab:xspec}
 }
\end{table}

\subsection{Comparison with prior archival data}
\label{sec:variability}

Detection of any flux variability can be important for understanding the nature of the source. The fluxes inferred from the two \c\ observations above are consistent with each other, and one must examine longer timescales. Searching in the \c\ archive for other data, one of the earliest ACIS-S observations covering the position of this source is sequence 600270 with observation date 2002 Jun 18. The exposure time after flare removal is only 18.03 ks, but there is a long time baseline of about 7 years to the 2009 observations analyzed in previous sections. 

\begin{figure}
  \begin{center}
    \includegraphics[height=8.5cm,angle=270]{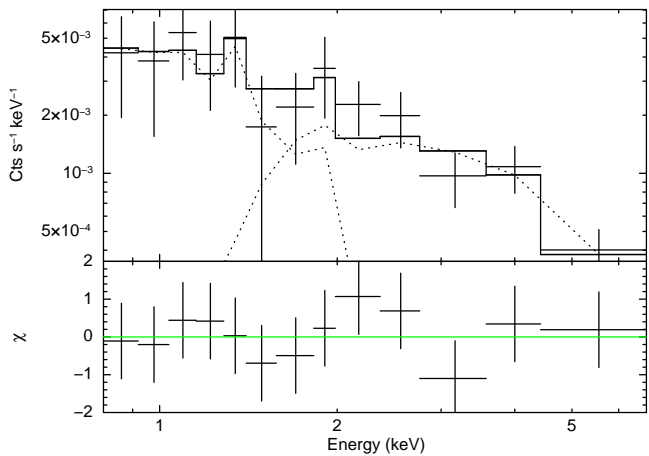}
  \end{center}
  \caption{ 
ACIS-S spectrum of J095552.7, extracted from an 18 ks exposure (sequence 600270) observed on 2002 Jun 18, seven years prior to the data shown in Fig.~\ref{fig:agn}, and fitted with the absorbed double thermal model described in \S~\ref{sec:xspec}. A minimum bin grouping of 20 counts has been applied.
\label{fig:2933} }
\end{figure}

We reduced and analyzed this archival data set in an identical manner to the 2009 observations. The extracted net spectrum is shown in Fig.~\ref{fig:2933}. Fitting with the same absorbed double thermal model as in \S~\ref{sec:xspec} yields parameters consistent with those listed in Table~\ref{tab:xspec} (though with much larger uncertainties), and a flux $F_{0.5-10}$=1.06$_{-0.2}^{+0.4}$$\times$10$^{-13}$ erg s$^{-1}$ cm$^{-2}$ corrected for Galactic absorption only. The 90\%\ confidence interval overlaps with the flux determined from the 2009 observations (\S~\ref{sec:xspec}). 

Thus, there is no indication for strong flux variability on a timescale of several years either. A similar inference has also been reached by \citet[][]{chiangkong11} in a recent analysis of other observations (their source designation is CXOU~J095552.8+694045). Our comparisons limit any flux variations to less than 20\%.

\section{Discussion}

We have presented the highest resolution imaging of the nuclear regions of M82 to date in the mid-IR. The Subaru diffraction limit at wavelengths of 11.7 and 12.81~\micron\ is 0\farcs 36 and 0\farcs 39, respectively. Compared to the previous best resolution observations (the 12.4~\micron\ imaging with PSF of 1$\farcs$1 presented by TG92), our images have PSFs improved by factors of 3.1 and 2.8 at 11.7 and 12.81~\micron, respectively, and we cover a core region larger by a factor of at least two. The referee also made us aware of a conference proceeding by \citet{ashby94}, where an image at 11.7~\micron\ with an angular resolution of 0\farcs 6 is presented. The absolute astrometry of their image agrees to within $\sim$1\arcsec\ with our work, as well as with the works of AL95 and TG92. To our knowledge, no further details have been published.

Multi-wavelength image comparisons show an anti-correlation between the observed stellar distribution (probed in the near-IR) with the distribution of warm dust that we probe. This means that obscuration in these dusty regions is heavy enough to scatter even near-IR radiation. 

Our observations provide the best view of the base of the dusty superwind that is known to exist in this galaxy, and reveal several elongated features on projected physical scales of up to $\ell$=120 pc at least. If the mid-IR emitting dust is mixed in with and entrained in outflowing gas, then the travel time ($t$) is 

\begin{equation}
t=5.9\left( \frac{\ell}{120\ {\rm pc}}\right) \left(\frac{200\ {\rm km\ s^{-1}}}{\rm v}\right) \times 10^5\ {\rm yr}.
\end{equation} 

\noindent
where a velocity of ${\rm v}$=200 km s$^{-1}$ identical to that found by \citet{nakai87} for the molecular gas is used as reference. This time period suggests recent energy input from young starbursts. The kinetic energy required to expel this dust is only a fraction of that channeled into the total gas mass, and so expulsion by SNe occurring at various locations around the ring of star-formation can easily account for this \citep{nakai87}. 

Using the integrated fluxes over the COMICS field-of-view alongside archival data at longer and shorter wavelengths, and assuming only a two temperature phase for the dust, we are able to describe the broad-band mid-IR diffuse emission with a $T$=160 K modified black body with a hot dust mass of $\sim$1000~\Msun, in addition to the 45 K cool dust component responsible for the far-IR emission. Assuming a standard gas:dust ratio of 100, the mass of the gas associated with the mid-IR emitting dust is then $\sim$10$^5$~\Msun. This is lower than the ionized gas mass of 2$\times$10$^6$~\Msun\ (TH80; \citealt{willner77}) by a factor of about 20, meaning that the ionized gas may be the predominant environment for the hot dust that we are observing. The cooler dust, on the other hand, is likely to be tracing the distribution of molecular gas (TH80). 

More than 20 discrete sources are detected, and most are found to have extended profiles. Matching against radio catalogs suggests that we have resolved at least four (and tentatively, five) \hii\ regions in the mid-IR for the first time. These \hii\ regions have monochromatic continuum dust luminosities ranging from $\nu L_{\nu}^{\rm 11.7\ \mu m}$=2$\times$10$^6$\lsun\ (radio ID 39.29+54.2) to 8$\times$10$^6$\lsun\ (41.17+56.2), consistent with being powered by embedded super star clusters similar to those seen in some other nearby galaxies \citep[e.g. ][]{galliano08}.

As seen in Fig.~\ref{fig:rgb}, no source is detected at the position of the radio transient identified by \citet{brunthaler09} as SN2008iz, the brightest radio SNR in M82 over the past 20 years. The start of bright radio flaring activity is limited to the time interval of 2007 Oct 29 to 2008 Mar 24. Our observations were carried out one month after the final set of follow-up radio observations on 2009 Apr 04 reported by \citeauthor{brunthaler09}, when they found the source to have an integrated 22.2 GHz flux of 9.2\p 0.2 mJy. Our flux limits are $F_{\rm NeII}$$<$38 mJy and $F_{\rm N11.7}$$<$18 mJy, assuming a point source of angular size equal to the diffraction limits in each filter. In general, there is little overlap between confirmed SNRs and mid-IR detections. Additionally, no mid-IR counterpart is detected at the location of the unusual radio transient found by \citet{muxlow10}. The source appeared on radio images obtained within the period of 1--5 May 2009 -- contemporaneous with our mid-IR observations -- and was not present one week before. It is located $\approx$1~\arcsec\ from the position of our source \# 18 (see also \citealt{kong09_atel} for an identification of the X-ray counterpart). We estimate point source upper limits of 34 mJy and 15 mJy in the NeII and N11.7 filters. 

No space observatory is foreseen to have a better resolving power than Subaru at $\sim$10~\micron, though the Mid InfraRed Instrument (MIRI) onboard \jwst\ should provide excellent sensitivity for only a modest loss in angular resolution at the same wavelengths \citep[e.g. ][]{jwstmiri}. MIRI is also expected to have a field of view larger by a factor of $\gtsim$2 on a side, making source identification much more secure. On a longer timescale, the European/JAXA mission \spica\ \citep{spica} will be crucial for deep searches of far-IR signatures of AGN activity at this source location. On the ground, the Gran Telecopio Canarias, with a primary mirror diameter of 10.4~m, can improve the resolution slightly to 0\farcs 3 under good observing conditions. Even better resolution must await larger ground-based observatories such as the Extremely Large Telescope. This will not be diffraction-limited under natural seeing conditions and will require additional adaptive optics capabilities in the mid-IR to improve upon the results presented herein.

\subsection{Nature of AGN candidate source}

The puzzling radio source 44.01+59.6 has a significant \neiifilter\ mid-IR detection (our source \# 18 [I52.70+45.9]), and is also coincident with a source visible in high-resolution X-ray images (J095552.7), to well within our estimated absolute positional uncertainties. From \c\ data, it is immediately apparent that the X-ray spectrum is not consistent with that of AGN, which usually display broad-band power-laws with photon-indices $\sim$1.9 characteristic of radiatively-efficient accretion \citep[e.g. ][]{mateos05_wide}. Low luminosity AGN with radiatively-inefficient flows or jets, too, usually display hard X-ray continua \citep[e.g. ][]{yu10}. Similarly, one can argue against direct association with other kinds of accreting sources such as X-ray binaries.

The detection of a strong He-like Fe line instead of a neutral line centered on 6.4 keV implies the absence of cold reflecting matter, also atypical for an accreting source. Highly ionized lines have actually been attributed to heavily obscured AGN in some cases \citep[e.g. ][ and references therein]{iwasawa05_arp220, nandraiwasawa07}, though on larger scales. In such a scenario, the AGN itself is not readily apparent; rather, the visible spectrum is dominated by surrounding gas which is irradiated by the AGN along directions out of our line of sight. If this is true for the case of source J095552.7, the underlying accreting object (be it an AGN or an X-ray binary) could power the radio jet found by \citet{wills99} and also provide a ready source where accreting gas is in rotation (as inferred from the maser lines found by \citealt{seaquist97}). But this would also result in X-ray photoionization spectral features or even an ionized reflection continuum. An extra layer of optically-thick obscuration is then also required (in addition to the two that we detect) with an extreme covering factor very close to unity so as to completely hide the reflection continuum and any neutral Fe K line. This is not consistent with the detection of a jet which would be expected to decrease the covering factor by clearing away some surrounding matter.

Finally, the 2--10 keV luminosity of 5$\times$10$^{38}$ erg s$^{-1}$ means that the source is unlikely to have a bolometric X-ray power which would place it in the ULX regime (e.g. \citealt{makishima00}) because of the steep spectra of the fitted thermal models.

Thus, our analysis allows us to reach some firm conclusions as to what the source is {\em not}. The true source nature still remains uncertain, though, and in the following, we discuss plausible alternatives.

In addition to our \neii\ image, the source appears most prominently as a compact object in hard X-rays. One possibility may then be that both the mid-IR and the hard X-ray components are associated with the SNR visible in the radio. To test this, we may compare the source power to that of Cas A, a powerful and young Galactic SNR. From \suzaku\ observations summed over the entire remnant, the 4--10 keV X-ray luminosity of Cas A is determined to be $L_{4-10\ {\rm keV}}^{\rm Cas A}$$\approx$4$\times$10$^{35}$ erg s$^{-1}$ \citep{maeda09}. This is about 150 times smaller than the deabsorbed X-ray luminosity $L_{4-10\ {\rm keV}}$$\approx$6$\times$10$^{37}$ erg s$^{-1}$ that we find for our source by using the high temperature {\sc apec} component alone (\S~\ref{sec:xspec}). This requires that the electron density ($n_e$) in the X-ray emitting interstellar medium (ISM) around source \# 18 be much larger than in the case of Cas A. $n_e$ may be determined by using the normalization values returned by the {\sc apec$_2$} component and the emission volume which, in the scenario under consideration here, should correspond to the volume of the radio-emitting plasma. \citet[][ see their Table 3]{fenech08} measure a diameter of 0.86~pc for the SNR, using which yields an electron density of $n_e$$\sim$1.75(\p 0.3)$\times$10$^3$ cm$^{-3}$. A low plasma filling factor would only push this value up. This is, indeed, much larger than typical ISM densities of 1--10 cm$^{-3}$ relevant for Galactic SNRs. We also note that the present limit on any X-ray variability in the source ($\ltsim$20\%\ over seven years) derived from archival data comparisons in \S~\ref{sec:variability} is consistent with the observed smaller flux changes in Cas A over similar timescales \citep[e.g. ][]{patnaude10}. The detected collimated radio jet may then suggest some kind of axisymmetry in the progenitor (e.g. a binary merger event).

On the other hand, the mid-IR power for source \# 18 ($\lambda L_{\lambda(12.81\ \mu{\rm m})}$=1.5$\times$10$^{40}$ erg s$^{-1}$) is in excess of the integrated 12~\micron\ \iras\ power of Cas A ($\lambda L_{\lambda(12\ \mu{\rm m})}^{\rm Cas A}$=5$\times$10$^{36}$ erg s$^{-1}$; \citealt{saken92}) by a much larger factor of $\approx$3000. But given the unknown dust temperature, the fraction of freshly formed vs. ambient heated dust, and the fact that our \neiifilter\ filter flux is likely to be dominated not by continuum but by the \neii\l12.81\micron\ line (which is known to be a strong cooling line for SNRs; e.g. \citealt{rho08}), further comparison of the mid-IR fluxes is difficult. 

\begin{figure}
  \begin{center}
    \includegraphics[height=8.5cm,angle=90]{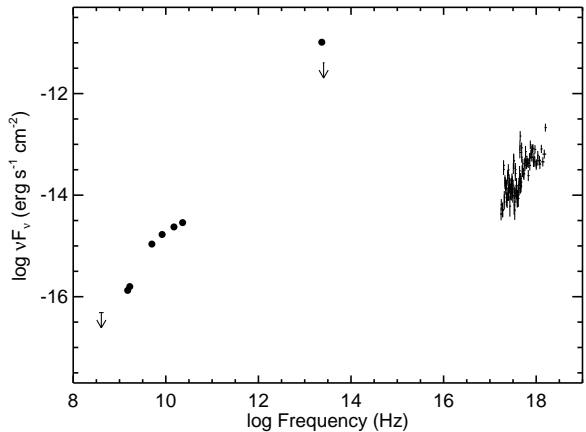}
  \end{center}
  \caption{ 
Broad-band SED of source \# 18. The radio data are the 1.3 cm to 74 cm fluxes (and one limit) from \citet{allenkronberg98}. In the mid-IR, we plot our \neiifilter\ filter flux, and a 3$\sigma$ \n117filter\ detection limit measured assuming Poisson statistics within an aperture equal in size to the diffraction limit. The X-ray regime shows the unfolded spectrum relevant for the model fitted to the observed data in Fig.~\ref{fig:agn}. 
\label{fig:nucleussed} }
\end{figure}

The other possibility to consider for the X-ray thermal plasma components is that of a hot ISM phase in a compact star cluster. In fact, collisionally ionized plasmas with a wide range of temperatures have been previously inferred to exist within starburst galaxies \citep[e.g. ][]{iwasawa05_arp220, ranalli08, strickland09}, though on larger scales. Fig.~\ref{fig:nucleussed} shows the compiled radio--to--X-ray fluxes for source \# 18. A mid-IR flux dominating the SED is consistent with an origin in a starburst, though this is not a unique solution. The detection of two distinct layers of absorption (\S~\ref{sec:xspec}) may also suggest a scenario which combines the above possibilities as follows. The hot X-ray thermal plasma and the radio counterpart could both be associated with a SNR, which is embedded within a host star cluster. The high column density affecting the {\sc apec}$_2$ component (Table~\ref{tab:xspec}) may easily be explained by strong local absorption within molecular clouds, for instance. The point-like profile of the hard X-ray data  is also consistent with this. The cluster, on the other hand, could appear prominently in soft X-rays and also in the mid-IR. The comparatively low aborption affecting this component is then due to gas along the plane of the galaxy on larger scales.

Further insight into the nature of this source will be possible if it can be isolated at longer wavelengths characterizing the peak of typical star-formation SEDs (this may be within reach of \jwst), or through detection of spectral features in the sub-mm with ALMA. Subsequent long \c\ exposures would be useful to place tighter constraints on X-ray variability. Meanwhile, the question of whether M82 hosts an AGN or not remains to be answered. 

\section{Acknowledgments}
PG acknowledges a JAXA International Top Young Fellowship and a RIKEN Foreign Postdoctoral Researcher Fellowship during parts of this work. This research is supported in part by a JSPS Grant in Aid Kakenhi number 21740152. PG also thanks S. Konami and P. Ranalli for discussions. The referee is acknowledged for their useful comments. 

    Based on data collected at Subaru Telescope and obtained from the SMOKA, which is operated by the Astronomy Data Center, National Astronomical Observatory of Japan. Data archives for the \c\ and \hst\ missions were essential for the multi-wavelength comparisons presented herein. Subaru observatory staff are thanked for their assistance in the observations. 

    This research has made use of SAOImage DS9 \citep{ds9}, developed by Smithsonian Astrophysical Observatory.

\label{lastpage}
\end{document}